\documentclass[final]{siamltex}
\pdfoutput=1
\usepackage{graphicx}
\title{Computing and deflating eigenvalues while solving multiple right hand 
	side linear systems with an application to Quantum Chromodynamics
\thanks{This work was partially supported by the National Science Foundation
grants ITR/DMR-0325218, CCF-0728915, and 
  the Jeffress Memorial Trust grant J-813.} }

\author{Andreas Stathopoulos \thanks{Department of Computer Science,
                College of William and Mary, Williamsburg,
                Virginia 23187-8795, ({\tt andreas@cs.wm.edu}).}
        \and
        Konstantinos Orginos \thanks{Department of Physics, 
                College of William and Mary, Williamsburg,
                ({\tt kostas@wm.edu}).}
	}

\begin{document}
\maketitle

\begin{abstract}
We present a new algorithm that computes eigenvalues and eigenvectors 
  of a Hermitian positive definite matrix while solving a linear system 
  of equations with Conjugate Gradient (CG). 
Traditionally, all the CG iteration vectors could be saved and recombined 
  through the eigenvectors of the tridiagonal projection matrix, 
  which is equivalent theoretically to unrestarted Lanczos.
Our algorithm capitalizes on the iteration vectors produced by CG
  to update only a small window of vectors that approximate the eigenvectors.
While this window is restarted in a locally optimal way, the CG algorithm 
  for the linear system is unaffected.
Yet, in all our experiments, this small window converges to the required 
  eigenvectors at a rate identical to unrestarted Lanczos.
After the solution of the linear system,
  eigenvectors that have not accurately converged can be improved 
  in an incremental fashion by solving additional linear systems.
In this case,
  eigenvectors identified in earlier systems can be used to deflate,
  and thus accelerate, the convergence of subsequent systems.

We have used this algorithm with excellent results in lattice QCD 
  applications, where hundreds of right hand sides may be needed.
Specifically, about 70 eigenvectors are obtained to full accuracy after 
  solving 24 right hand sides. 
Deflating these from the large number of subsequent right hand sides 
  removes the dreaded critical slowdown, 
  where the conditioning of the matrix increases as the quark mass reaches
  a critical value.
Our experiments show almost a constant number of iterations for our method, 
  regardless of quark mass, and speedups of 8 over original CG 
  for light quark masses.
\end{abstract}


\section{Introduction}

The numerical solution of linear systems of equations of large, 
  sparse matrices is central to many scientific and engineering applications. 
One of the most computationally demanding applications is
 lattice Quantum Chromodynamics (QCD) because not only does it involve 
  very large matrix sizes but also requires the solution of several
  linear systems with the same matrix but different right hand sides.
Direct methods, although attractive for multiple right hand sides, 
  cannot be used because of the size of the matrix.
Iterative methods provide the only means for solving these problems.

QCD is the theory of the fundamental force known as strong interaction, 
  which describes the interactions among quarks, one of the constituents of 
  matter.
Lattice QCD is the tool for non-perturbative numerical calculations
 of these interactions on 
  a Euclidean space-time lattice \cite{Wilson:1974sk}.
The heart of the computations is the solution of the lattice-Dirac equation,
 which translates to a linear system of equations 
	$Mx = b$,
 often for a large number of right hand sides \cite{Foley:2005ac}.
The Dirac operator $M$ is $\gamma_5$-Hermitian,
  or $\gamma_5 M = M^H \gamma_5$, where, in one representation, $\gamma_5$ 
  is a diagonal matrix with 1 and -1 on the diagonal.
Also, $M = m_q I- D$, where $m_{q}$ is a parameter related to the quark mass 
  and $D$ is an operator.
In addition to solving linear systems of equations, many current approaches 
  \cite{Foley:2005ac,Edwards:2001ei,DeGrand:2004qw,Giusti:2004yp}
  require the solution of the eigenvalue problem 
  $(\gamma_5 M)u_i = \lambda_i u_i$,
  for 100-200 smallest magnitude eigenvalues, $\lambda_i$, 
  and their eigenvectors, $u_i$ (together we call them eigenpairs).
Beyond the very large dimension and number of right hand sides, $M$ 
   becomes increasingly ill-conditioned as $m_{q} \rightarrow m_{critical}$.
In lattice QCD this is known as critical slowdown and is a limiting 
   computational factor.

Traditionally these linear systems are solved by applying the 
  Conjugate Gradient (CG) on the $A = M^HM$ Hermitian operator.
Although there are cases where BICGSTAB on the nonsymmetric operator 
  can be twice as fast 
	\cite{Frommer_bicgstabdQCD,Fischer:1996th}, in other cases,
  such as in domain wall fermions, CG is not only the fastest but 
  also the most robust method\footnote{Private communication Tom Blum and Taku Izubuchi.}.
In this paper, we focus on CG for Hermitian systems for two reasons.
  First, CG is characterized by optimal convergence.
  Second, the eigenvectors of $A$ are also eigenvectors of $\gamma_5 M$.
  Hence, after an initial CG phase, computed eigenvectors can be used to 
  deflate BICGSTAB on $\gamma_5 M$ for the rest of the right hand sides.

Solving a linear system with many right hand sides is 
  an outstanding problem.
Traditional approaches solve each system one by one.
Unknowingly, such methods regenerate search directions within 
  previously explored subspaces, thus wasting iterations in
  successive right hand sides.
Sharing information between systems has long been recognized as 
  the key idea \cite{CLanczos_bcg}. 
This is typically performed through seed methods 
  \cite{VS_EG_95,TFC_WLW_94,Saad_87a} 
  or through block methods 
  \cite{Golub_Underwood_77,DOLeary_80a,Guennouni_Jbilou_Sadok}.
For Hermitian matrices, a selective sharing of only the useful part of 
  information between systems can be achieved through invariant subspaces.
Such ideas have been tried in QCD \cite{deForcrand:Progress}, 
  but effective deflation methods have only appeared recently.
In section \ref{sec:Background} we review deflation methods 
  that either precompute the required eigenpairs or
  use a restarted method to compute eigenvectors 
  while solving linear systems.
Both approaches, however, are unnecessarily expensive.

In this paper we present an algorithm that computes eigenvalue and eigenvector
  approximations of a Hermitian operator by reusing information from the 
  {\em unrestarted} CG method.
This is achieved by keeping a search space that includes current eigenvector 
  approximations and only the last few CG iteration vectors.
The crucial step is how we restart this search space to keep computations
  tractable.
The CG iteration is completely unaffected.
Our experiments show that eigenvector convergence is similar to 
  unrestarted Lanczos; an impressive achievement yet to be understood 
  theoretically.

Our motivating application is the computation of nucleon-nucleon scattering
   \break
  lengths with all to all propagators, where, for a time discretization 
  between 64 and 128, 1500-3000 right hand sides must be solved.
Our algorithms are equally applicable to the more classic problem of 
  computing nucleon form factors using sequential propagators where 
  120 right hand sides are required.
In a first phase, we solve 24 right hand sides with our method.
Unconverged eigenvectors from one system improve incrementally during the 
  solution of subsequent systems.
In the second phase, after 24 right hand sides, enough eigenvectors have 
  been obtained to significantly reduce the condition number of the 
  deflated matrix in the classic CG.
In this phase we observe speedups of 8-9 over the non-deflated CG, and, 
  most importantly, the number of the deflated CG iterations remains almost constant 
  as $m_{q}$ approaches $m_{critical}$, thus removing the critical slowdown.

\section{Background and current approaches}
\label{sec:Background}

Krylov methods provide one of the most important tools for solving large, 
  general sparse linear systems.
An excellent survey of recent developments and discussion on some of the 
  open problems in linear systems appears in 
  \cite{Valeria_Szyld_RecentDevelopments}.
For symmetric (or Hermitian) positive definite (SPD) matrices, CG 
  \cite{MH_ES_CG} remains the uncontested choice because it uses a 
  three term recurrence to converge optimally, with minimum storage 
  and computational requirements. 
Even for the Hermitian case, however, it remains an open question as to 
  how best solve a system with many (say $s$) right hand sides, 
			$$Ax_i = b_i,\ i=1,\ldots ,s.$$

One approach is to use block methods which work simultaneously on a set 
  of vectors \cite{Golub_Underwood_77,DOLeary_80a,Guennouni_Jbilou_Sadok}.
They have favorable performance characteristics in memory hierarchy computers 
  and usually reduce the number of iterations.
However, their implementation is involved as linear dependencies 
  in the block must be removed \cite{blockQMR}.
More importantly, the total execution time often increases and there is 
  no clear theoretical understanding of when to use them and with how 
  large a block size (see 
   \cite{Gutknecht_website_block,Gutknecht_block_intro} for a recent review
   and \cite{Morgan_Wilcox} for a discussion of block methods in the context
   of multiple right hand sides and QCD).

The other common approach is the use of seed methods 
  \cite{VS_EG_95,TFC_WLW_94,Saad_87a}, which reuse 
  the Krylov subspace generated by one seed system to project the rest.
After projection, a new seed system is chosen to iterate to convergence, 
  and the idea is repeated until all systems are solved.
Seed methods effectively reduce the number of iterations to solve 
  each successive right hand side when these are highly related.
Most seed algorithms do not store the previously generated Krylov spaces.
Instead, while solving $Ax_i = b_i$, they project the current Krylov 
  vector from
  all the approximations $x_j, j=i+1,\ldots ,s$, updating all remaining linear 
  systems at every iteration \cite{VS_EG_95,TFC_WLW_94}.
Seed and block methods have also been combined \cite{Misha_seed}.
Still, this type of seed methods presents three difficulties:
First, the $b_i$ vectors may not be all available at the same time.
Second, we consider the contribution of each Krylov vector to all systems.
This contribution is usually too small to warrant the additional expense,
  so the total time may increase.
Third, the second problem becomes extreme when the $b_i$ are unrelated.

The above difficulties can be avoided by noticing that for unrelated $b_i$ 
  and for SPD matrices the only improvements from 
  seed methods should come almost entirely from those common subspaces 
  that a Krylov method builds for any starting vector: 
  the extreme invariant subspaces.
In particular, the eigenvalues near zero should be targeted as their
  deflation dramatically decreases the condition number of the matrix.
Also, we would like to avoid using an eigensolver to compute those eigenpairs
  but to reuse the Krylov space built by CG.

For non-Hermitian matrices, the GMRESDR method 
  \cite{RBMorgan_2002,Morgan_Wilcox} computes approximate eigenspace 
  information during GMRES($m$) \cite{GMRES}.
In GMRESDR, when the GMRES basis reaches $m$ vectors, it is restarted not only 
  with the residual of the approximate solution of the linear system but also 
  with the $nev < m$ smallest magnitude Ritz pairs.
It is known that GMRESDR is equivalent to the Implicitly Restarted 
  Arnoldi \cite{Sorensen_92}, so while GMRESDR solves the linear system,
  the $nev$ Ritz pairs also converge to the smallest magnitude eigenpairs.
This elegant solution transfers also to the Hermitian case, where GMRESDR
  becomes equivalent to thick restarted Lanczos \cite{TRLAN,AS_YS_KW_98}.

For Hermitian systems, however, the GMRESDR approach presents three
  disadvantages. 
First, it is based on GMRES which is much more expensive per step than CG.
Second, restarting GMRES every $m$ steps impairs the optimal convergence of CG.
Third, restarting also impairs the convergence of unrestarted Lanczos to
  the $nev$ required eigenpairs.
In the context of this paper, the latter is an important disadvantage,
  because the eigenspaces may not be obtained at the accuracy 
  required for deflation of other systems.
In that case, we would like to incrementally improve on the approximate 
  eigenspace during the solution of subsequent linear systems.
This is performed in \cite{Giraud_incremental}, but only for GMRESDR.
In Section \ref{sec:ourmethod} we present a way to compute eigenpairs 
  from the {\em unrestarted} CG.
One of the components of our method is similar to a method developed
  independently by Wang et al. \cite{Wang_Sturler_Paulino}.
As we show later, our method is not only cheaper but by making the appropriate
  restarting choices it yields practically optimal convergence.

\subsection{Deflating the CG}
\label{sec:back_initCG}
Once a set of seed vectors $U$ have been computed, there are several ways 
  to use $U$ to ``deflate'' subsequent CG runs.
Assume that the solution $x$ of the system $Ax = b$ has significant 
  components in $U$.
Given an initial guess $\tilde x_0$, we can consider another guess whose 
  $U$ components are removed ahead of time.
This is performed by the Galerkin oblique projection \cite{Saad_linear}:
\begin{eqnarray}
\label{eq:initCG}
x_0 = \tilde x_0 + U(U^HAU)^{-1}U^H(b-A\tilde x_0).
\end{eqnarray}
Then $x_0$ is passed as initial guess to CG.
These two steps are often called \mbox{init-CG}
   \cite{TFC_WLW_94,Giraud_Ruiz_Touhami_06,initCG}.
The init-CG approach works well when $U$ approximates relatively 
  accurately the eigenvectors with eigenvalues closest to zero.

When $U$ spans an approximate eigenspace that has been computed to 
  $\epsilon$ accuracy,
  init-CG converges similarly to deflated CG up to $\epsilon$ accuracy 
  for the linear system. 
After that point, convergence plateaus and eventually becomes similar 
  to the original CG \cite{Giraud_Ruiz_Touhami_06}.
A variety of techniques can be employed to solve this problem:
  using $U$ as a ``spectral preconditioner''
  \cite{Erhel_precg_mrhs,Nicolaides_deflation,Giraud_spectral,Brower:1995vx}, 
  including $U$ as an explicit orthogonalization constraint in CG
  \cite{Kaasschieter,Wang_Sturler_Paulino}, or combinations of methods
  as reviewed in \cite{Giraud_Ruiz_Touhami_06}.
Typically, these techniques result in convergence which is almost identical 
  to exact deflation, but the cost per iteration increases significantly.
In this paper, we focus only on the init-CG method.

\subsection{Locally optimal restarting for eigensolvers} 

We conclude this background section with a restarting technique for 
  eigensolvers that plays a central role in the method we develop in this paper.
Because the Hermitian eigenvalue problem can be considered a constrained 
  quadratic minimization problem, many eigenvalue methods have been developed
  as variants of the non-linear Conjugate Gradient (NLCG) method 
  on the Grassman manifold \cite{EdelmanAriasSmith}.
However, it is natural to consider a method that minimizes the
  Rayleigh quotient on the whole space
  $\left\{ u^{(m-1)}, u^{(m)}, g^{(m)} \right\},$ 
  instead of only along one search direction.
By $\theta^{(m)}, u^{(m)}$ we denote the eigenvalue and eigenvector 
  approximations at the $m$th step and 
  $g^{(m)} = Au^{(m)} - \theta^{(m)} u^{(m)}$ the corresponding residual.
The method:
\begin{eqnarray}
  u^{(m+1)} & = & 
   \mbox{RayleighRitz}\left(\{ u^{(m-1)}, u^{(m)}, g^{(m)} \}\right),\ m>1,
  \label{eq:locg}
\end{eqnarray}
  is often called locally optimal Conjugate Gradient (LOCG)
  \cite{DYAKONOV_83,Knyazev_91}, and seems to consistently outperform
  other NLCG type methods.
For numerical stability, the basis can be kept orthonormal,
  or $u^{(m)}-\tau^{(m)} u^{(m-1)}$ can be used instead of $u^{(m-1)}$,
  for some weight $\tau^{(m)}$.
The latter is the LOBPCG method \cite{LOBPCG}.
  
Quasi-Newton methods use the NLCG vector iterates to construct incrementally
  an approximation to the Hessian, thus accelerating NLCG 
  \cite{Gill_Murray_Wright}.
Similarly, if all the iterates of LOCG are considered, 
  certain forms of quasi-Newton methods are equivalent to unrestarted Lanczos.
With thick or implicit restarting, Lanczos loses this 
  single important direction ($u^{(m-1)}$).
Therefore, the appropriate way to restart Lanczos or Lanczos-type methods 
  is by subspace acceleration of the LOCG recurrence 
  \cite{CM_SR_ED_92a,AS_YS_98}.
When looking for $nev$ eigenvalues, the idea can be combined with thick 
  restarting so that the eigensolver search basis is
  restarted with an orthonormal basis for the following vectors
  \cite{AS_YS_98,JDQMR_One}:
\begin{eqnarray}
\left[ 
 u_1^{(m)}, u_2^{(m)}, \ldots , u_{nev}^{(m)}, u_1^{(m-1)}, \ldots ,u_k^{(m-1)}
\right].
  \label{eq:restartedV}
\end{eqnarray}
An efficient implementation is possible at no additional cost to thick
  restarting, because all orthogonalization is performed on the small 
  coefficient vectors of the Rayleigh-Ritz procedure \cite{AS_YS_98}.
This technique consistently yields convergence which is almost 
  indistinguishable from the unrestarted method.
We see next how this scheme can help us approximate eigenvectors from within CG.

\section{The eigCG method}
\label{sec:ourmethod}

The idea is to use an eigenvector search space $V$ within the CG iteration, 
  which is restarted through (\ref{eq:restartedV}),
  but not to let it dictate the next search direction.
Instead, we leverage a window of the last $m$ residuals computed by the 
  unrestarted CG to build an appropriately restarted subspace.

Consider the preconditioned CG algorithm in the left box of 
  Figure \ref{fig:eigCG}.
We will exploit the equivalence of CG and Lanczos and the fact that 
  the Lanczos vectors are the appropriately normalized CG 
  residuals \cite{GolubVanLoan}.
In the general case of an SPD preconditioner, $P\neq I$, 
  the Lanczos method approximates eigenpairs of the symmetric matrix 
  $\hat A = P^{-1/2}AP^{-1/2}$ from the Lanczos basis
  $P^{-1/2} R = P^{-1/2}[r_1/\rho_1,\ldots ,r_j/\rho_j]$,
  where $\rho_j = (r_j^Hz_j)^{1/2}$ and $z_j = P^{-1/2}r_j$.
Let $\hat U = P^{-1/2}R Y$ and $\hat \Lambda$ be the Lanczos approximations
  to the eigenvectors and eigenvalues of $\hat A$, respectively.
These are exactly the eigenpairs needed to deflate subsequent linear systems 
  with $\hat A$ using (\ref{eq:initCG}).
The inconvenient $\hat U$ basis can be avoided because CG needs an
  initial guess $x_0 \approx x$ and not $\hat x_0 \approx P^{1/2}x$.
Specifically, let $V = P^{-1}R = [z_1/\rho_1, \ldots ,z_j/\rho_j]$, 
  let $\tilde x_0$ be some approximation to $x$, 
  and $P^{1/2}\tilde x_0$ the corresponding initial guess for the 
  split preconditioned system $\hat A$.
If $\hat x_0$ is the deflated initial guess for system $\hat A$ 
  by applying (\ref{eq:initCG}) on $P^{1/2}\tilde x_0$,
  then the initial guess $x_0$ to be given to CG is:
\begin{eqnarray*}
x_0 =  P^{-1/2} \hat x_0 
    &=& P^{-1/2}(P^{1/2}\tilde x_0 + \hat U \hat \Lambda^{-1} \hat U^H P^{-1/2} 
					  (P^{-1/2}b-P^{-1/2}A\tilde x_0)) \\
    &=& \tilde x_0 + P^{-1} RY \hat \Lambda^{-1} Y^HR^H P^{-1} 
							(b - A\tilde x_0)\\ 
    &=& \tilde x_0 + VY \hat \Lambda^{-1} Y^HV^H (b-A\tilde x_0).
\end{eqnarray*}
The $Y$ are the eigenvectors corresponding to the eigenvalues $\hat \Lambda$
  of the Lanczos tridiagonal matrix 
  $T_m = V^H A V = (P^{-1/2}R)^H \hat A (P^{-1/2}R)$.
The $T_m$ is obtained at no extra cost from the scalars
  $\alpha_j = r_{j-1}^Hz_j / (p_j^HAp_j)$ and 
  $\beta_j = r^H_{j-1}z_{j}/(r_{j-2}^Hz_{j-1})$ 
  computed during the CG iteration \cite[p.194]{Saad_linear}:
\begin{eqnarray}
\hspace{12pt}
T_m = \left[ \begin{array}{ccccc}
 1/\alpha_1	      &  \sqrt{\beta_2}/\alpha_1\\
\sqrt{\beta_2}/\alpha_1 & 1/\alpha_2 +\beta_2/\alpha_1 \\
& & \ddots & & \sqrt{\beta_{m+1}}/\alpha_m\\
& & & \sqrt{\beta_{m+1}}/\alpha_m & 
				1/\alpha_m + \beta_m/\alpha_{m-1} \\
	\end{array}
	\right].
\label{eq:tridiagonal}
\end{eqnarray}

Traditionally, the Lanczos Ritz values and vectors are computed 
  from $T_m$ at the end of CG.
This, however, requires the storage of all $z_j$ 
  or a second CG run to recompute them on the fly, effectively doubling 
  the cost.
Moreover, if the number of CG iterations is large, dealing with 
  spurious eigenvalues at the end is expensive. 
Instead, we introduce an algorithm that restarts the search space for 
  computing eigenvalues but does not restart CG.

The proposed algorithm, eigCG, adds new functionality to CG
  as shown in Figure \ref{fig:eigCG} in a Matlab-like format.
It uses a set of $m$ vectors, $V$, to keep track of the 
  lowest $nev$ Ritz vectors (or more accurately the $P^{-1/2}\hat U$) 
  of the matrix $\hat A$.
Initially, $V$ is composed of the normalized CG preconditioned residuals 
  $z/\sqrt{r^Hz}$ (step 11.13),
  and the projection matrix $T_m = V^HAV$ is available from
  (\ref{eq:tridiagonal}) (steps 11.1 and 11.10).
When $V$ reaches $m$ vectors, we restart it exactly as we would 
  restart an eigensolver using (\ref{eq:restartedV}) with $k=nev$.
The Rayleigh Ritz can be applied both on $T_m$ and $T_{m-1}$ to produce 
  Ritz vector coefficients for the last two consecutive steps 
  (step 11.3).
We also append a zero row to $\bar Y$ (step 11.4), so that 
  $Y$ and $\bar Y$ have the same dimension $m$. 
In steps 11.5, 11.6 we use the Rayleigh Ritz again to compute an 
  orthonormal Ritz basis for the space spanned by $[Y, \bar Y]$.
In step 11.7 the restarted basis $V$ and its corresponding diagonal 
  projection matrix are computed.
Notice $V$ remains orthonormal (or $P$-orthonormal in case of preconditioning).
After restarting, eigCG continues to append the CG preconditioned residuals 
  $z$ to $V$, and to extend $T_i$ with the same tridiagonal coefficients.
The only exception is the first residual added after restart.
It requires a set of inner products, $z^HAV$, to explicitly update the 
  $i+1$ row of matrix $T$.
As shown in step 11.8, a matrix vector product can be avoided if 
  we remember the previous vector $t_{prev}$ (step 10.1) and note that
  $Az = Ap - \beta_j t_{prev}$.
After this point, no other extra computation is needed until the next restart
  ($m-2nev$ iterations).

\begin{figure}[h]
{\normalsize
\begin{minipage}[t]{.35\textwidth}
\fbox{
\begin{minipage}[t]{\textwidth}
\begin{tabbing}
x\=xxx\=xx\=xx\=\kill
{\bf The CG algorithm}\\
\> 0 \> $r = b - Ax$; $j = 0$   \\
\> 1 \> while $\|r\|/\|r_0\| > tol$    \\
\> 2 \>\> $j = j + 1$   \\
\> 3 \>\> $z = P^{-1}r$    \\
\> 4 \>\> $\rho_{prev} = \rho; \ \rho = r^Hz$   \\
\> 5 \>\> if ($j$ == 1) \\
\> 6 \>\>\>     $p = z$ \\
\> 7 \>\> else   \\
\> 8 \>\>\>     $\beta_j = \rho/\rho_{prev}$    \\
\> 9 \>\>\>     $p = z + \beta_j p$       \\
\> 10 \>\> end   \\
\> 11 \>\> $t = Ap$   \\
\> 12 \>\> $\alpha_j = r^Hz/(p^Ht)$   \\
\> 13 \>\> $x = x + \alpha_j p$   \\
\> 14 \>\> $r = r - \alpha_j t$   \\
\> 15 \> end   
\end{tabbing}
\end{minipage}
}
\end{minipage}
\fbox{
\begin{minipage}[t]{.55\textwidth}
\begin{tabbing}
x\=xxxxxx\=xx\=xx\=\kill
{\bf The eigCG($nev,m$) additions to CG}\\
\> 0.1  \> $V = [\ ]$; $i = 0$  \\
\> 10.1  \> if $(i == m)$, $t_{prev} = t$\\
\> 11.1 \> if $(j>1),\ T_{j-1,j-1} = \frac{1}{\alpha_{j-1}} 
		+ \frac{\beta_{j-1}}{a_{j-2}}$\\
\> 11.2 \> if $(i == m)$ \\
\> 11.3 \>\>   Solve for $nev$ lowest eigenpairs of \\
\>      \>\>\>   $T_m Y = Y M$ and $T_{m-1} \bar Y = \bar Y \bar M$\\
\> 11.4 \>\>   Add an $m$th row of zeros: $\bar Y = [\bar Y ; 0]$ \\
\> 11.5 \>\>   $Q$=orth($[Y \bar Y]$), set $H = Q^H T_m Q$\\
\> 11.6 \>\>   Solve for the eigenpairs of $H Z = Z M$\\
\> 11.7 \>\>   Restart:\\
\>      \>\>\>  $V = V (Q Z)$, $i = 2nev$, $T_{i} = M$ \\
\> 11.8 \>\>   Compute $z_j^HAV$, the $i+1$ row of $T_{i+1}$:\\ 
\>      \>\>\>  $w = t - \beta_j t_{prev}$ \\
\>      \>\>\>  $T_{i+1,1:i} = w^HV/\sqrt{\rho}$\\
\> 11.9 \> else \\
\> 11.10\>\>   $T_{i+1,i} = -\sqrt{\beta_j}/\alpha_{j-1}$\\
\> 11.11\> end \\
\> 11.12\> $i = i+1$\\
\> 11.13\> $V(:,i) = z/\sqrt{\rho}$ 
\end{tabbing}
\end{minipage}
} 
} 
\caption{The eigCG($nev,m$) algorithm approximates $nev$ eigenvalues keeping 
  a search space of size $m > 2nev$.
The classical CG algorithm is shown in the left box. 
To obtain the eigCG algorithm we extend the CG steps 0, 10, and 11,
  with new steps numbered with decimal points.
The right box shows these additions to CG.
}
\label{fig:eigCG}
\end{figure}

Computationally, eigCG requires storage for $m$ vectors, but 
  no additional matrix-vector operations or other costs during 
  the CG iterations except at restart.
At restart, all operations that involve matrices of size $m$, 
  which includes the orthogonalization at step 11.5, have negligible cost.
The only two expenses are: step 11.7 which requires $O(2N\cdot m\cdot 2nev)$ 
  flops for $V (QZ)$, 
  and step 11.8 which requires $O(2N\cdot 2nev)$ flops for $z_j^HAV$. 
This expense occurs every $(m-2nev)$ iterations, for a total of 
   $$O(4N\cdot nev\cdot (m+1)/(m-2nev)) \mbox{ flops per iteration.}$$
Interestingly, the average cost per step decreases with $m$
  and for large enough $m > 10nev$ it approaches $O(4 N nev)$ flops per step.
This is the same as the computation of $2nev$ Ritz vectors from 
  the full basis of unrestarted Lanczos. 
Finally, the extra computations are fully parallelizable:
  First, the computation $V (QZ)$ incurs no synchronization and can be 
  performed in a cache efficient way with level 3 BLAS routines.
  Second, the accumulation of the dot products $V^HAz$ can be delayed 
  until the first dot product during the next CG iteration to avoid an 
  extra synchronization.

\subsection{Convergence and comparison with other methods}
\label{sec:convergence}

In exact arithmetic, the vectors in $V$ remain orthonormal 
  (or $P$-orthonormal in case of preconditioning) as linear 
  combinations of the Lanczos vectors.
In floating point arithmetic, the Lanczos property guarantees 
  $z_j \perp V$ and the numerical accuracy of the $T_m$ coefficients 
  until some eigenpairs converge to square root of 
  machine precision ($\sqrt{\epsilon_{mach}}\|A\|$).
After that, spurious eigenvalues start to appear in $T_{m}$ but
  without compromising the accuracy of the correct ones.
Converging but unwanted eigenpairs, lying in the high end of the spectrum, 
  pose little stability threat because they are purged at every 
  restart (step 11.7).
Nevertheless, we should choose $m$ such that the highest eigenpair 
  does not converge in $m-2nev$ iterations, i.e., between successive restarts.
Unless the largest eigenvalues are highly separated, this is not 
  usually an issue.

To avoid spurious wanted eigenvalues,
  our eigCG algorithm facilitates a straightforward implementation of 
  selective orthogonalization and a cheaper version of 
  partial orthogonalization \cite{SEP}.
For any Ritz vector $Vy$, where $y$ is any column of $Y$,
  its residual norm can be monitored at no additional cost through 
  the Lanczos property, $\| AVy - \mu_i Vy\| = |\beta_j||y_{m}|$.
When $V$ is restarted we can selectively orthogonalize against 
  all those Ritz vectors whose accuracy approaches $\sqrt{\epsilon_{mach}}$.
Alternatively, we can simply orthogonalize the CG residual against 
  all the restarted $V$.
The additional expense is minimal and the benefits are twofold; 
  spurious eigenvalues are avoided in $T_m$ and 
  the convergence of CG improves as well.
We do not further explore this approach in this paper for three reasons;
First, to keep the presentation simple and focused on the main eigCG idea;
Second, as shown in section \ref{sec:incremental}, we use eigCG incrementally 
  which resolves spurious issues at a higher level;
Third, in our QCD application code we use single precision for all 
  computations except for dot products. Therefore we do not expect any 
  spurious eigenvalues to show up while solving one linear system.

The structure of $T_m$ after restart 
  is reminiscent of the thick restarted Lanczos (TRLAN) \cite{TRLAN}.
However, after the first restart, the residual of TRLAN (equivalently 
  of ARPACK \cite{arpack} or of GMRESDR \cite{RBMorgan_2002}) 
  is the residual produced by $2nev$ steps of 
  {\em some other Lanczos process with a different initial vector.
Our residuals continue to be the Lanczos vectors of the original CG/Lanczos 
  process.}
Therefore, the convergence of the CG is unaffected.
Also, eigenvalue convergence in eigCG is expected to be different 
  from TRLAN or GMRESDR.

Recently, the recycled MINRES (RMINRES) algorithm was developed independently 
  in \cite{Wang_Sturler_Paulino}.
It is based on the same technique of reusing some of the MINRES residuals
  in a basis $V$.
The critical difference with eigCG is that RMINRES restarts as in TRLAN
  by keeping only the $nev$ harmonic Ritz vectors closest to zero
  and not the previous directions (i.e., $k=0$ in (\ref{eq:restartedV})).
As a result, {\em eigenvalues converge only to a very low 
  accuracy and then stagnate}.
This is acceptable in \cite{Wang_Sturler_Paulino} because their application
  involves systems of slightly varying matrices which cannot be deflated
  exactly by each other's eigenvectors.
Moreover, RMINRES tries to identify and maintain directions that tend 
  to repeat across Krylov subspaces of different linear systems. 
To be effective, the basis $V$ must be kept orthogonal to 
  these vectors, thus increasing the expense of its MINRES iteration.
Our eigCG focuses on getting the eigenvectors accurately which can later
  be deflated inexpensively with init-CG.

The use of Rayleigh Ritz and thick restart guarantee monotonic convergence
  of the $nev$ Ritz values from $V$.
Beyond that, it is not obvious why eigCG should converge to any eigenpairs, 
  let alone with a convergence rate identical to that of unrestarted Lanczos!
With thick restarting alone (as in RMINRES) the important information that 
  is discarded cannot be reintroduced in $V$ as future residuals of the 
  unrestarted CG are orthogonal to it.
By using the restarting of eq.~(\ref{eq:restartedV}), with modest $nev$ values,
  almost all Lanczos information regarding the $nev$ eigenpairs is kept 
  in condensed form in $V$.
Then, the new CG residuals do in fact represent the steepest descent 
  for our Ritz vectors, and eigCG behaves like unrestarted Lanczos.

\begin{figure}
\includegraphics[width=.5\textwidth]{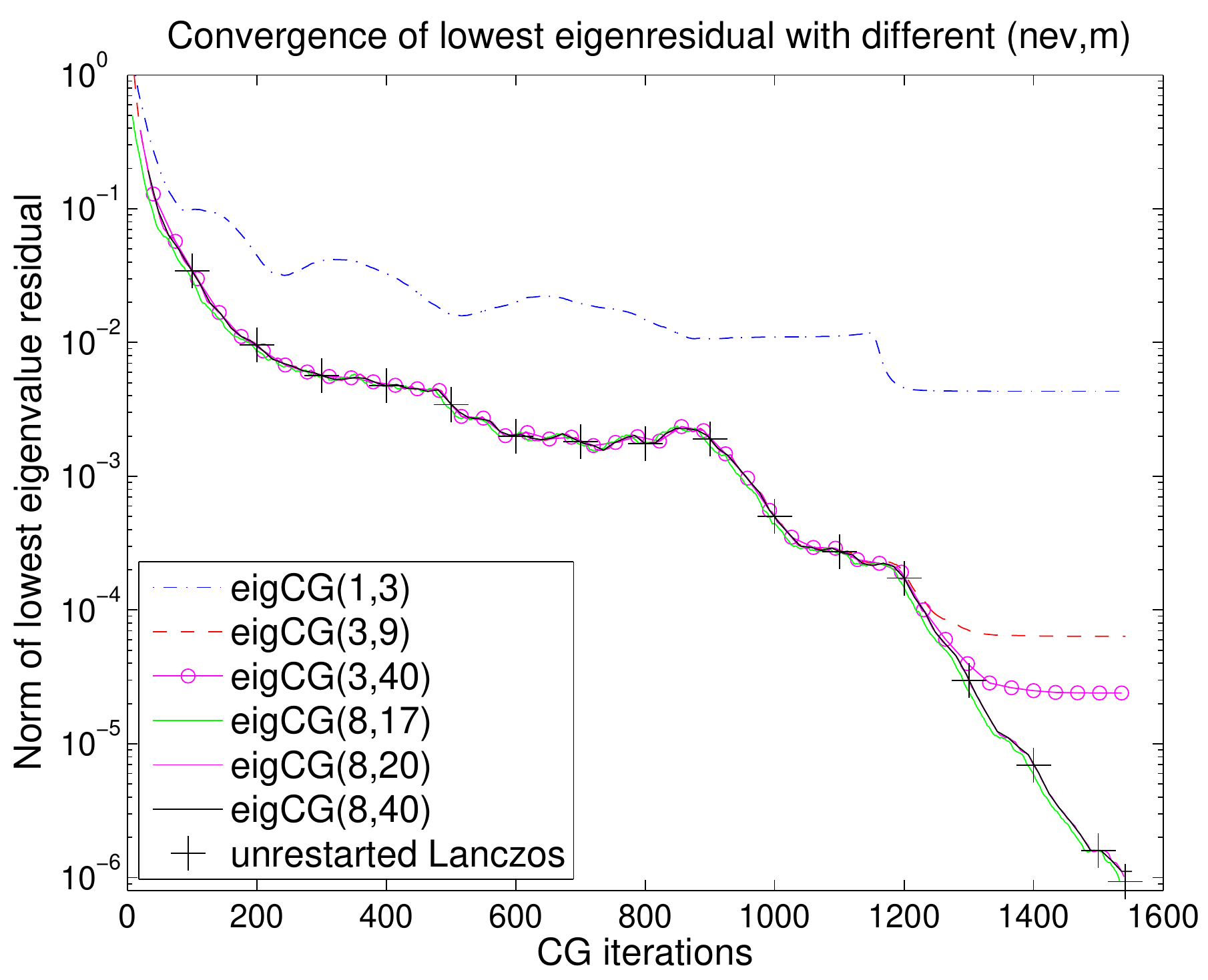}
\includegraphics[width=.5\textwidth]{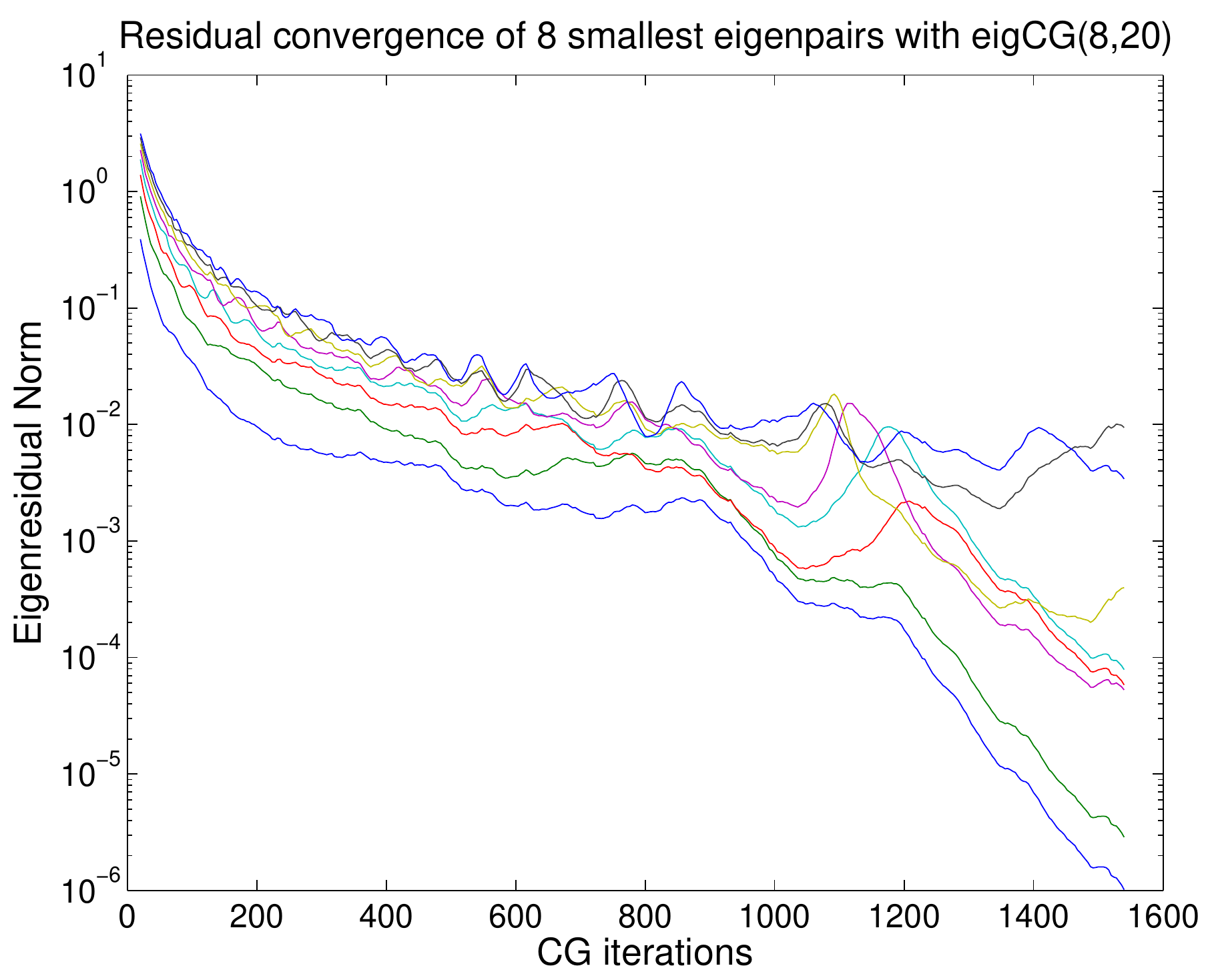}

\caption{Left graph: eigCG convergence with various values of ($nev,m$), 
   	matching unrestarted Lanczos.
	Right graph: convergence of 8 smallest eigenpairs of eigCG(8,20).}
\label{fig:eigCG_conv}
\end{figure}

The left graph of Figure \ref{fig:eigCG_conv} shows residual convergence
  for the smallest eigenpair under various eigCG($nev,m$) runs on a matrix 
  of size $12\times 8^4 = 49152$ that represents the spectrum of a typical 
  Wilson fermion matrix with light quark mass.
The eigCG(1,3) holds only three vectors in $V$, which cannot capture 
  the information well, and stagnates.
Keeping as few as three Ritz vectors (eigCG(3,9)) improves the stagnation 
  point dramatically,
  while with eight Ritz vectors we were able to reach accuracy of 1e-12. 
The figure only shows convergence until step 1540 which is where
  the linear system converged.
It is remarkable, however, that until it reaches the stagnation point eigCG
  converges at the rate of unrestarted, fully reorthogonalized Lanczos.
We have observed the same property for all $8$ smallest eigenvalues,
  whose convergence is
   depicted in the right graph in Figure \ref{fig:eigCG_conv}.
A theoretical analysis of this surprising behavior will be the focus 
  of our future research.

\begin{figure}

\includegraphics[width=\textwidth]{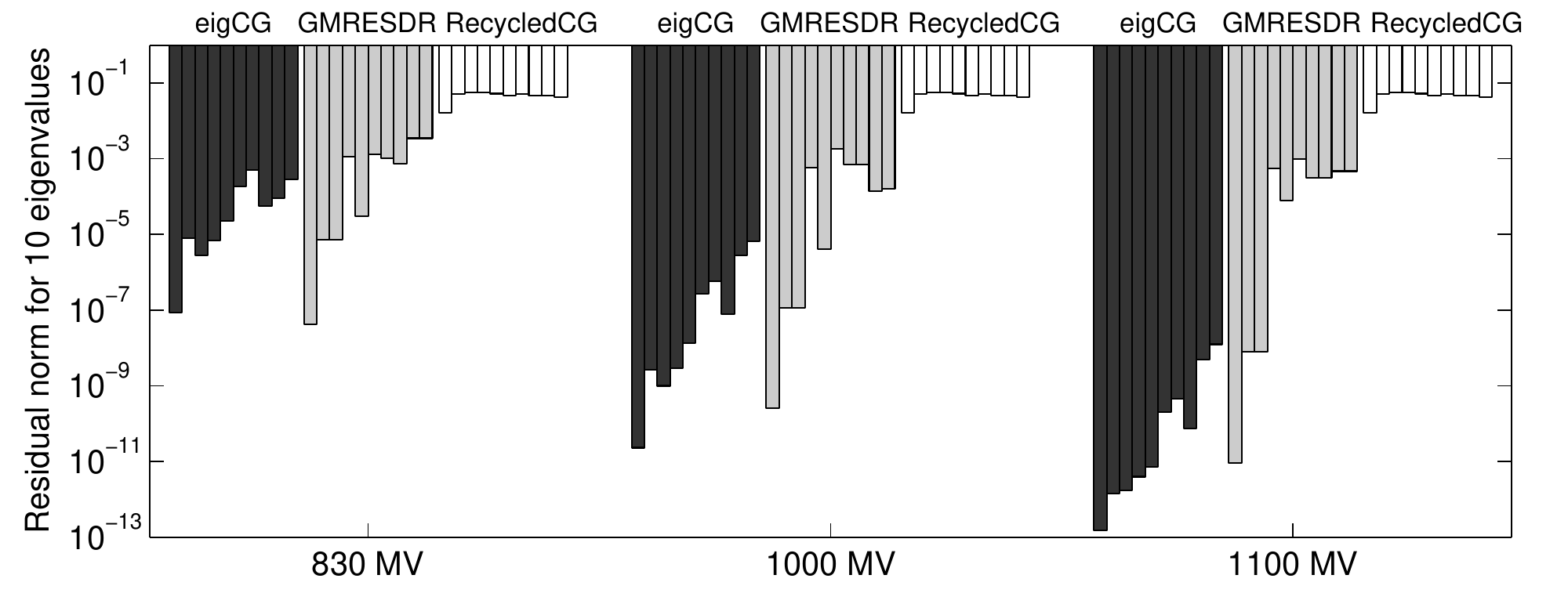}
\caption{Accuracy of 10 smallest eigenvalues computed by eigCG(10,50), 
  GMRESDR(55,34), and RecycledCG(50,30) methods, at 830, 1000, and 
  1100 iterations. For each method and case, 
  smallest eigenvalues correspond to leftmost bars.
}
\label{fig:comparisons}
\end{figure}

Figure \ref{fig:comparisons} compares the effectiveness of eigCG in 
  approximating many eigenpairs to those of the competing GMRESDR and 
  RecycledCG methods (the latter implementing the RMINRES ideas on CG).
The experiments are run on a Wilson Fermion lattice of size $12\times 12^4$
  with periodic boundary conditions and quark mass equal to the critical mass.
We use odd-even preconditioning (which yields a matrix of half the size)
  and focus on the symmetric normal equations.
We run the experiments in Matlab 7.5, on a Mac Pro, dual processor, 
  dual-core, 3 GHz Intel Xeon.
We test eigCG(10,50), GMRESDR(55,34), and RecycledCG(50,30)
  so that all methods have similar memory requirements.
We plot the residual norms of the 10 smallest eigenvalues computed by the
  three methods at 830, 1000, and 1100 matrix-vector products.
GMRESDR is effective at computing the lowest three eigenvalues but the accuracy
  of the rest does not seem to improve with further iterations.
RecycledCG cannot improve the low accuracy obtained early during the iteration.
Our method not only computes more accurate and more eigenvalues than both 
  other methods, it also continues to improve them with more iterations.
Moreover, GMRESDR took 1168 iterations and 1505 seconds to solve the 
  linear system to accuracy $10^{-8}$, while eigCG took 1010 iterations and
  410 seconds.
Thus, eigCG was 3.7 times faster, while also computing a much better 
  eigenspace.

\begin{figure}
\centering
\includegraphics[width=0.8\textwidth]{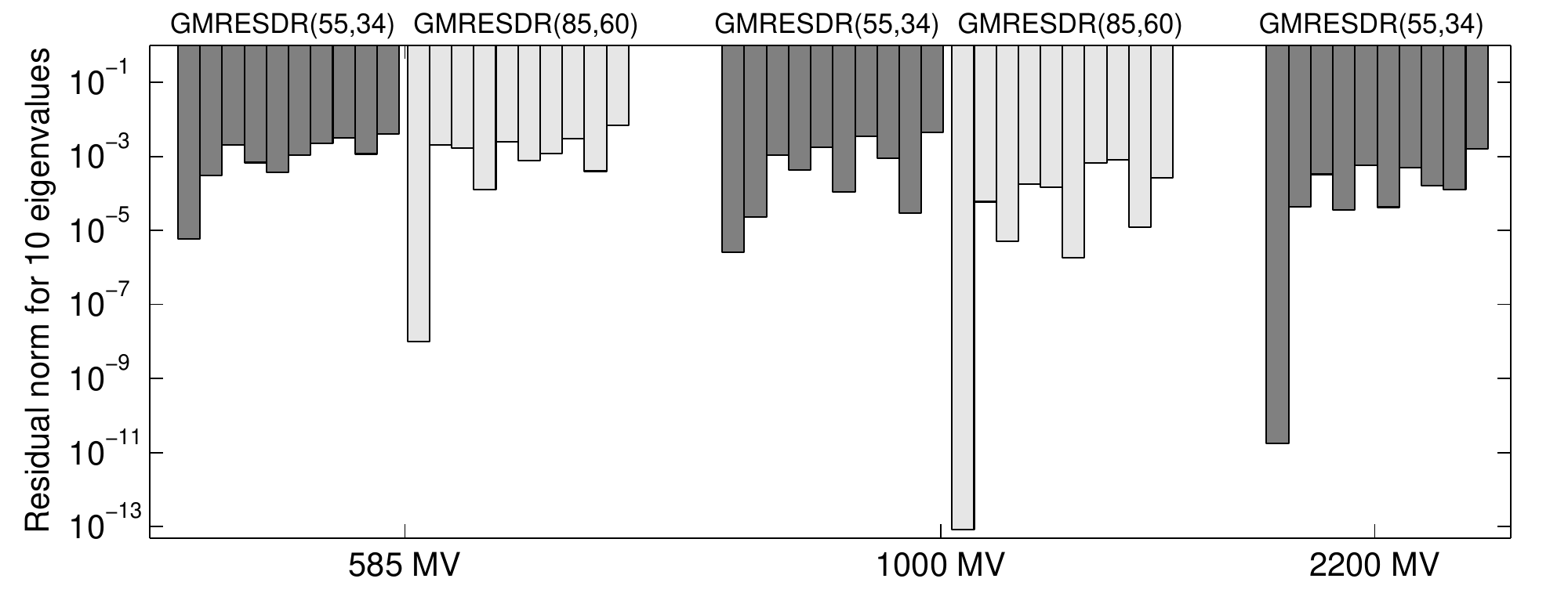}
\caption{Accuracy of 10 smallest magnitude eigenvalues computed by 
  GMRESDR(55,34) and GMRESDR(85,60), at 585, 1000, and  2200 iterations
  on the nonsymmetric operator.
  GMRESDR(85,60) is not reported at 2200 MV since it reached machine 
  precision much earlier.
}
\label{fig:GMRESDR}
\end{figure}

For comparison, we also report some results from running the GMRESDR 
  directly on the nonsymmetric Dirac operator.
This is sometimes reported to be twice as fast as CG on the normal 
  equations because it avoids the squaring of the condition number. 
However, such results are often obtained from heavier quark masses,
  where the problem is not as difficult or interesting.
In our $12\times12^4$ problem, and with the same accuracy and 
  computational platform, GMRESDR(55,34) took 1000 iterations and 1120 seconds.
GMRESDR(85,60) improved convergence to 585 iterations, but its execution
  time was still 1001 seconds.
For low quark masses, as in our case, GMRESDR is not only slower than eigCG,
  but it has more difficulty computing accurate eigenpairs even with more 
  memory as shown in Figure~\ref{fig:GMRESDR}.

\section{Incrementally increasing eigenvector accuracy and number}
\label{sec:incremental}
When the CG iteration converges before equally accurate eigenvectors
  can be obtained, these eigenvectors cannot deflate effectively the CG for 
  the next right hand side.
Deflating with the eigenvectors obtained in the example of 
  Figure \ref{fig:eigCG_conv} yielded 10\% faster convergence in the 
  next linear system.
As we mentioned earlier, using the resulting $V$ as a spectral
  preconditioner is much more effective, resulting 
  in 50\% faster convergence in the same example.
Still, we would like to avoid this expense on every step of CG, 
  and more importantly to obtain more eigenvectors of $A$ for more effective
  deflation.

A simple outer scheme can be designed that calls eigCG for $s_1 < s$ 
  right hand sides and accumulates the resulting $nev$ approximate Ritz vectors
  from each run into a larger set, $U$.
\begin{figure}[h]
\fbox{
\begin{minipage}{\textwidth}
\begin{tabbing}
xxxx\= \mbox{\hspace{2.62in}} \=\kill
{\bf The Incremental eigCG algorithm}\\
$U = [\ ], H = [\ ]$      	    \>\> \% accumulated Ritz vectors\\
for $i$ = 1 : $s_1$   		    \>\> \% for $s_1$ initial rhs \\
\>  $x_0 = U H^{-1} U^H b_i$        \>   \% the init-CG part \\
\>  [$x_i, V, M$ ] = eigCG($nev, m, A, x_0, b_i$)
				    \> \% eigCG with initial guess $x_0$ \\
\>  $\bar V =$ orthonormalize $V$ against $U$ 
				    \> \% (Not strictly necessary) \\
\>  $W = A\bar V$,
    $H = \left[ \begin{array}{cc}       H & U^HW \\
				    W^HU  & \bar V^HW 
		\end{array} \right]$
				    \> \% Add $nev$ rows to $H$\\
\>  Set $U = [U,\ \bar V]$ 	    \> \% Augment $U$\\
end
\end{tabbing}
\end{minipage}
}
\end{figure}

\vspace{-11pt}
\noindent
If we assume that $U$ contains converged eigenvectors, the Ritz vectors 
  produced in $V$ during the solution of the next linear system will be 
  in the orthogonal complement of span$(U)$, as $x_0$ is deflated of $U$.
If some of the vectors in $U$ are not converged enough, the eigCG 
  produces $V$ with directions not only in new eigenvectors but also 
  in directions that complement the unconverged Ritz vectors in $U$.
Thus, the accuracy of $U$ incrementally improves up to machine precision
  as our experiments indicate.

Although it is advisable for computational reasons to keep $m$ large, 
  $nev$ should not be too large because it increases the eigCG cost 
  without providing more than a few good eigenpair 
  approximations from one linear system.
Computing many eigenpairs through the Incremental eigCG allows us 
  to choose modest values for $nev$. 
In our experiments, we used $nev=10$ while larger values
  did not consistently improve the results.
Other dynamic techniques for setting $nev$ based on monitoring eigenvalue
  separation and convergence are also possible but not 
  explored in this paper.

Computationally, if the number of vectors in $U$ is $l$, 
  the above algorithm costs $nev$ matrix-vector operations 
  and $(4 l nev + 4 nev^2)N + 2 l nev N$ flops of level 3 BLAS
  for each new system solved.
For very large $l$ this time could become noticeable, in which case
  we could use the following two optimizations.
First, we note that orthogonalization of the new vectors is only performed
  to reduce the conditioning of $H$, so it does not need to be carried
  out accurately.
Moreover, it would suffice to orthogonalize only those Ritz vectors in $V$ 
  with large components in $U$.
These can be identified by whether their Ritz values lie within the range 
  of already computed eigenvalues.
Second, we could lock accurate eigenvectors out of $U$ so that they do not
  participate in building the $H$ matrix.
Naturally, if the number of right hand sides is large enough, $s \gg s_1$, 
  any of the above costs are amortized by the much faster convergence of the 
  deflated systems for $b_i$,  $i > s_1$.
Nevertheless, our timings show that the overhead of the incremental part 
  is small even for 240 vectors.
  
Storage for the total of $l = s_1 nev$ vectors of $U$ is usually not a challenge
  because it is on the order of the right hand sides or the solutions
  for the $s$ linear systems (assuming $s \geq s_1 nev$). 
Moreover, $U$ is not accessed within eigCG or CG, so it can be kept in 
  secondary storage if needed.
Still, it would be beneficial computationally and storage-wise
  if we could limit the size of $U$ to only a certain number of
  smallest eigenvectors.
We have observed that if after augmenting $U$ with $V$,  
  we truncate $U$ to include only a certain number of smallest eigenvectors, 
  the accuracy ceases to increase with $V$ from subsequent systems.
Although this problem merits further investigation, in the QCD problems
  in which we are currently interested the number of right hand sides
  is large enough to allow us to grow $U$ to a large number.

\section{Numerical experiments}
In this section we present results from two real world applications.
After describing the computational platform and applications,
  we address four issues.
In section \ref{sec:numerical-manyevals},
  we study convergence of eigenvalues in the Incremental eigCG.
In section \ref{sec:numerical-initcgworks},
  we study the improvement in convergence of the linear 
  systems as bigger spaces are deflated in init-CG. 
In section \ref{sec:numerical-noslowdown}, 
  we study the convergence invariance of the resulting
  deflated init-CG under various quark masses.
Finally, in section \ref{sec:numerical-timings},
  we provide timings that show the small overhead and high efficiency 
  of the method on a supercomputing platform.

\subsection{Chroma implementation and two lattice QCD tests}
We implemented our algorithms in C and interfaced with Chroma, 
  a lattice QCD C++ software base developed at 
  Jefferson National Lab~\cite{Edwards:2004sx}.  
All our experiments were done in single precision with dot product
  summations in double precision.
In the following three sections, we report tests on an 8 node dual socket, 
  dual core cluster with 4GB of memory per node.
The timings in section \ref{sec:numerical-timings} are reported
  from a production all-to-all propagator calculation on 256 processors 
  of the Cray XT4 at NERSC, Lawerence Berkeley National Lab.

The Dirac matrices used in this paper come from two ensembles of 
  an anisotropic, two flavor dynamical Wilson fermion calculation.
In both cases the sea pion mass is about 400(36) MeV and the lattice 
  spacing is 0.108(7)fm. 
The anisotropy factor of the time direction is about 3 and the 
  critical mass was determined to be -0.4188 in all cases. 
We studied the behavior of our algorithm on several different lattices 
  from these ensembles and found insignificant variation in performance.
In one case the lattice size is $16^3 \times 64$ 
  for a matrix size of $N=3,145,728$, 
  while in the other it is $24^3\times 64$ for a matrix size of 
  $N= 10,616,832$.
We refer to these cases as 3M and 10M lattices, respectively.
The odd-even preconditioned normal equations were solved in both cases
  with storage requirements about $100N$ per matrix.
These ensembles are used in current lattice QCD calculations 
  with more than one hundred right hand sides 
  at Jefferson Lab, hence our algorithmic improvements have direct 
  implications in the cost of currently pursued lattice QCD projects.

\begin{figure}
\includegraphics[width=0.5\textwidth]{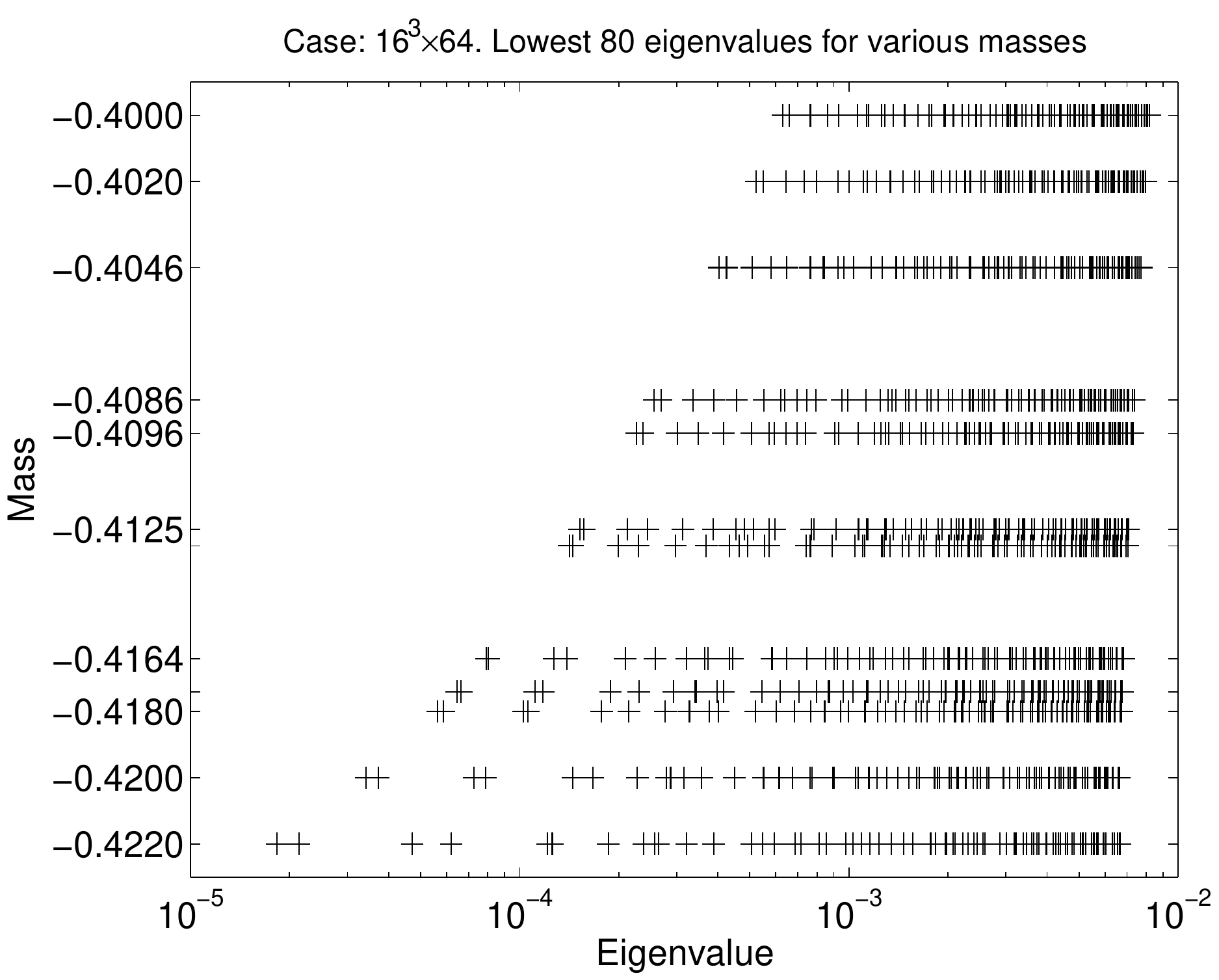}
\includegraphics[width=0.5\textwidth]{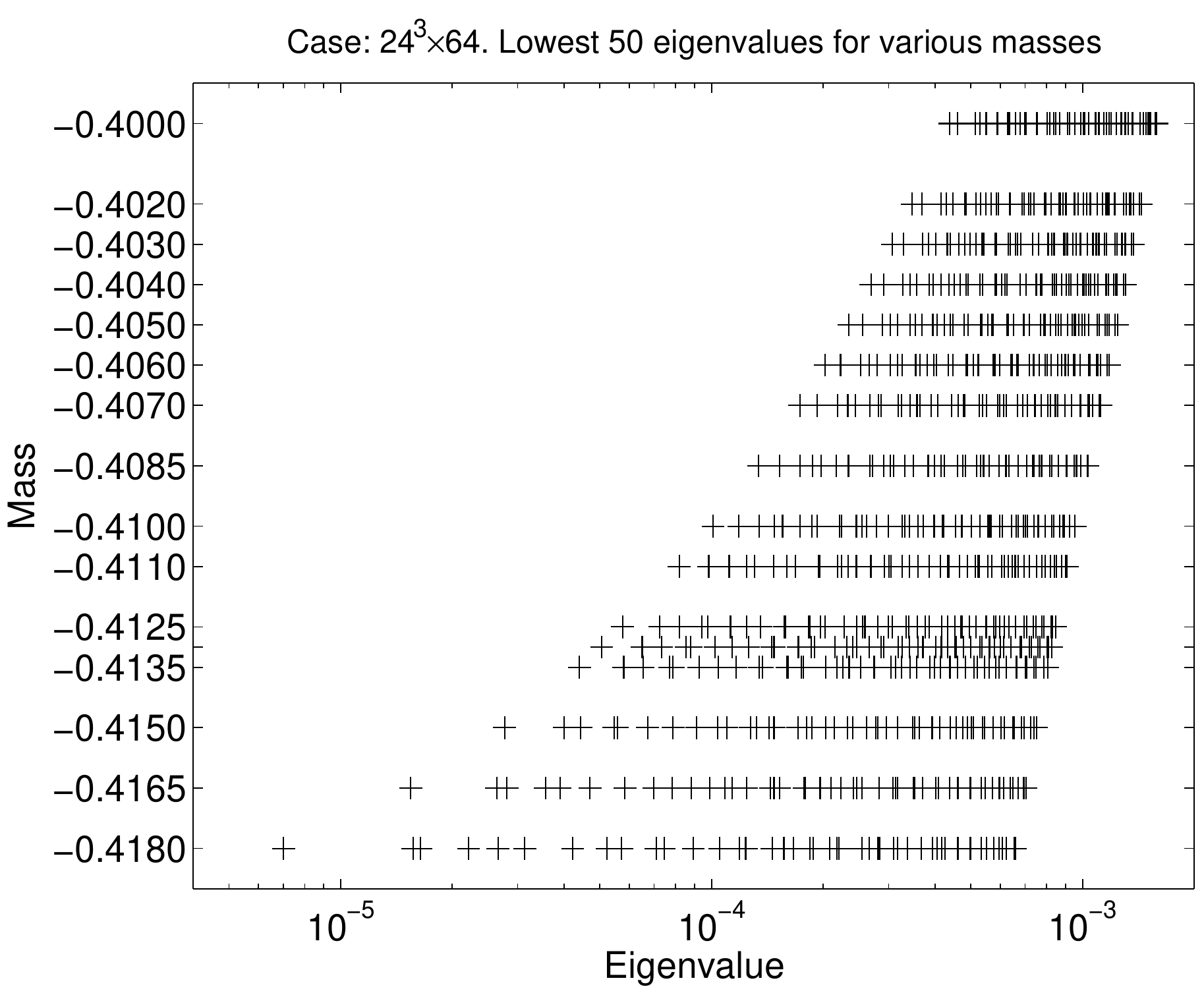}
\caption{The smallest eigenvalues for the $M^+M$ positive definite matrix under
  various quark masses. 80 smallest eigenvalues are shown for the 3M lattice
  (left graph) and 50 smallest eigenvalues for the 10M lattice (right graph).}
\label{fig:eigenvalues}
\end{figure}

In Figure \ref{fig:eigenvalues} we present the lowest part of the spectrum 
  for the matrices resulting from the two test lattices and for a range of 
  quark masses. 
As expected, the lowest eigenvalue becomes smaller as the quark 
  mass approaches the critical quark mass (-0.4188)
  leading to large condition numbers and to the critical slowdown.
However, the more interior an eigenvalue is the lower the rate 
  that it decreases with the mass.
This also is expected, because the largest eigenvalue and the average 
  density of the spectrum is a function of the discretization volume and 
  not as much of the quark mass  \cite{Spectral_Gaps_in_QCD}. 
Therefore, by deflating a sufficient number of lowest eigenvalues not only
  do we reduce the condition number significantly, but we also make it 
  practically constant regardless of quark mass, thus removing the 
  critical slowdown.
  
For the 3M lattice, for example, deflating 12 vectors from the lightest masses
  yields a condition number similar to the heavy mass case -0.4000.
For the 10M lattice, about 30 vectors are required for the same effect. 
These examples show the limitation of all deflation methods, not only eigCG. 
As eigenvalue density increases with volume, more eigenvalues need to be
  deflated to achieve a constant condition number.
Traditionally, scalability across volumes is the topic of multigrid methods
  \cite{Brandt_77,Brandt_86,Brower:1991ni,Brannick:thesis,Brannick:2006}.
Although, our methods are very effective in that direction too, 
  they only guarantee constant time across different masses.

\subsection{Eigenvalue convergence in Incremental eigCG}
\label{sec:numerical-manyevals}

\begin{figure}
\includegraphics[width=0.5\textwidth]{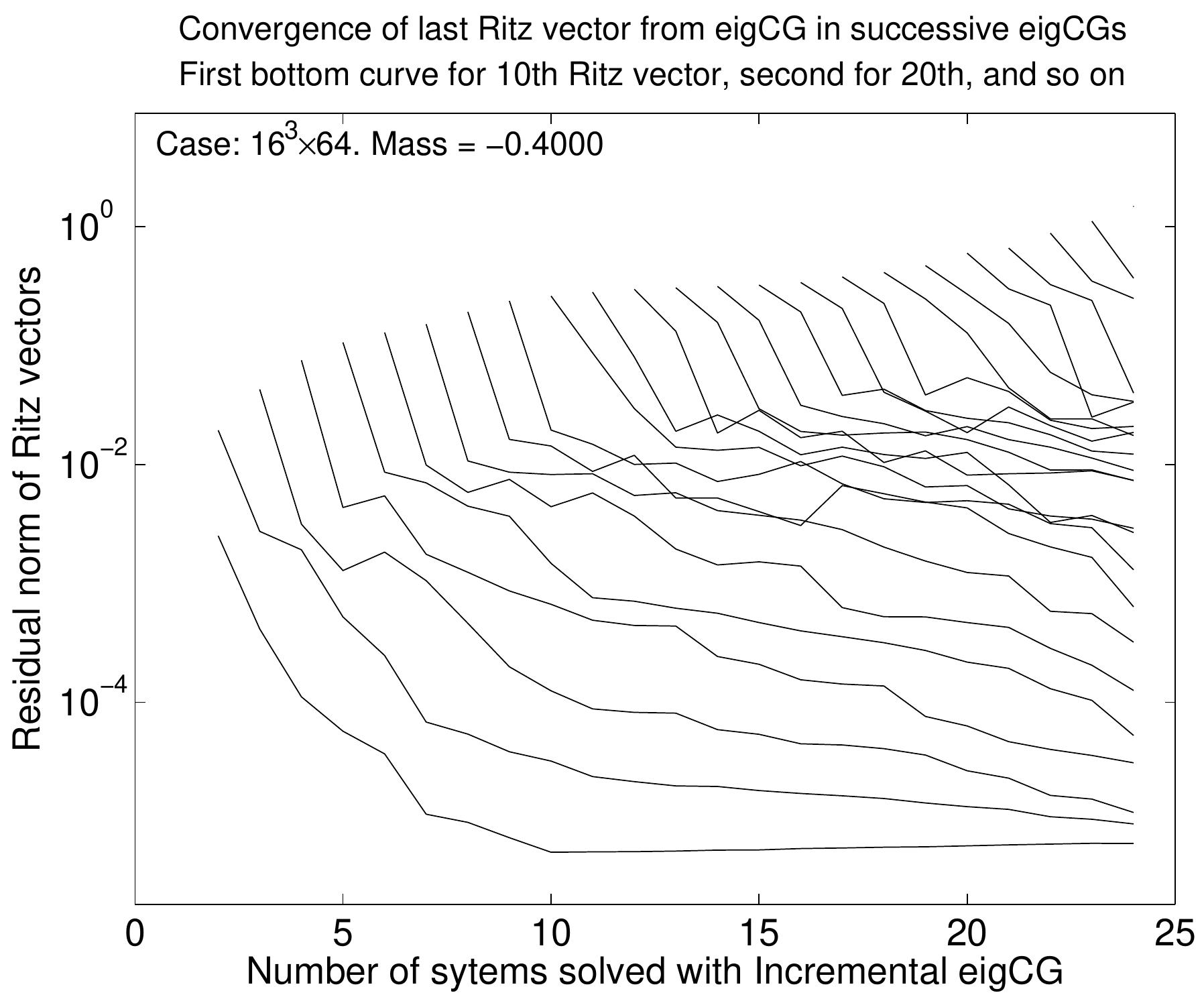}
\includegraphics[width=0.5\textwidth]{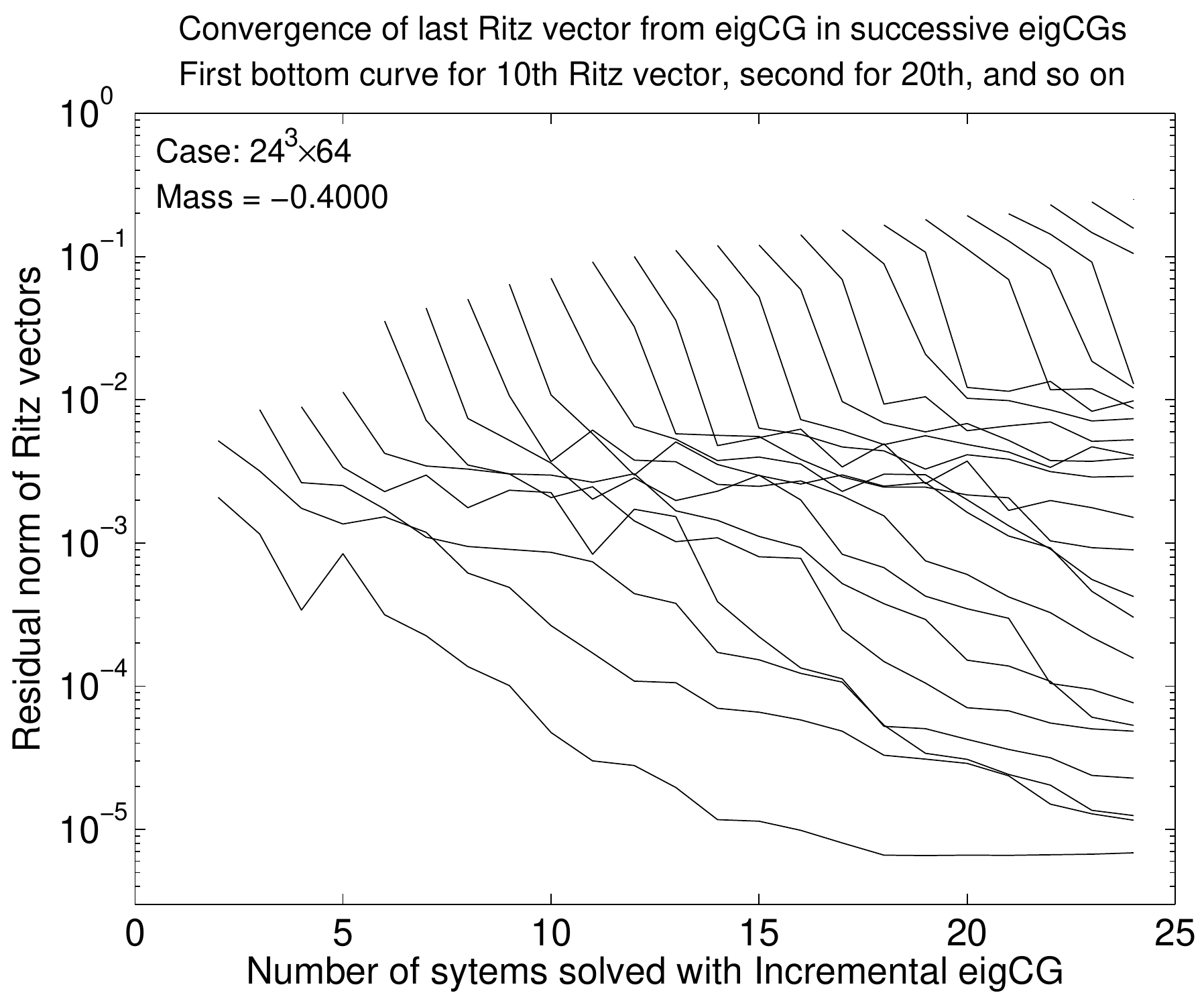}

\includegraphics[width=0.5\textwidth]{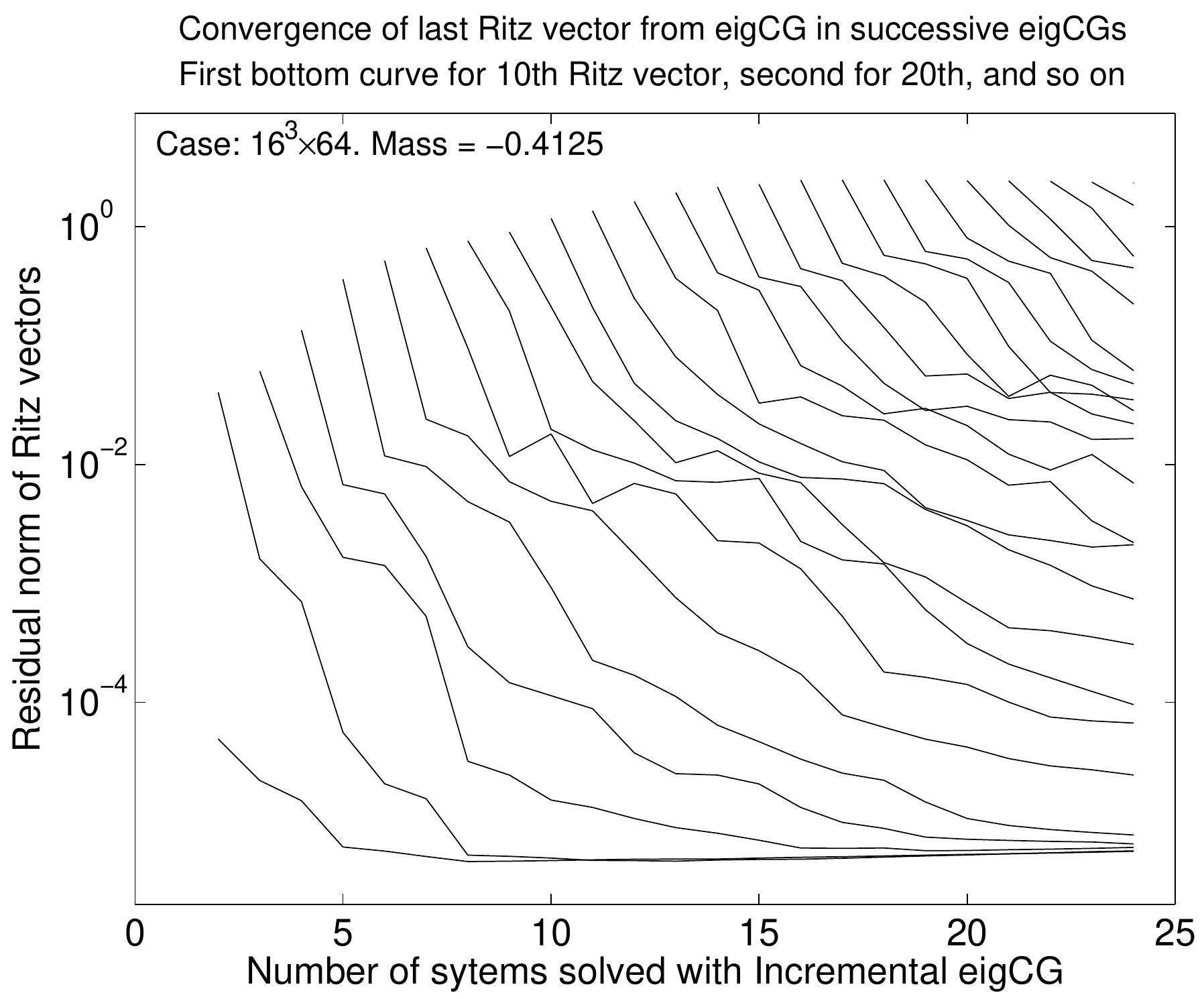}
\includegraphics[width=0.5\textwidth]{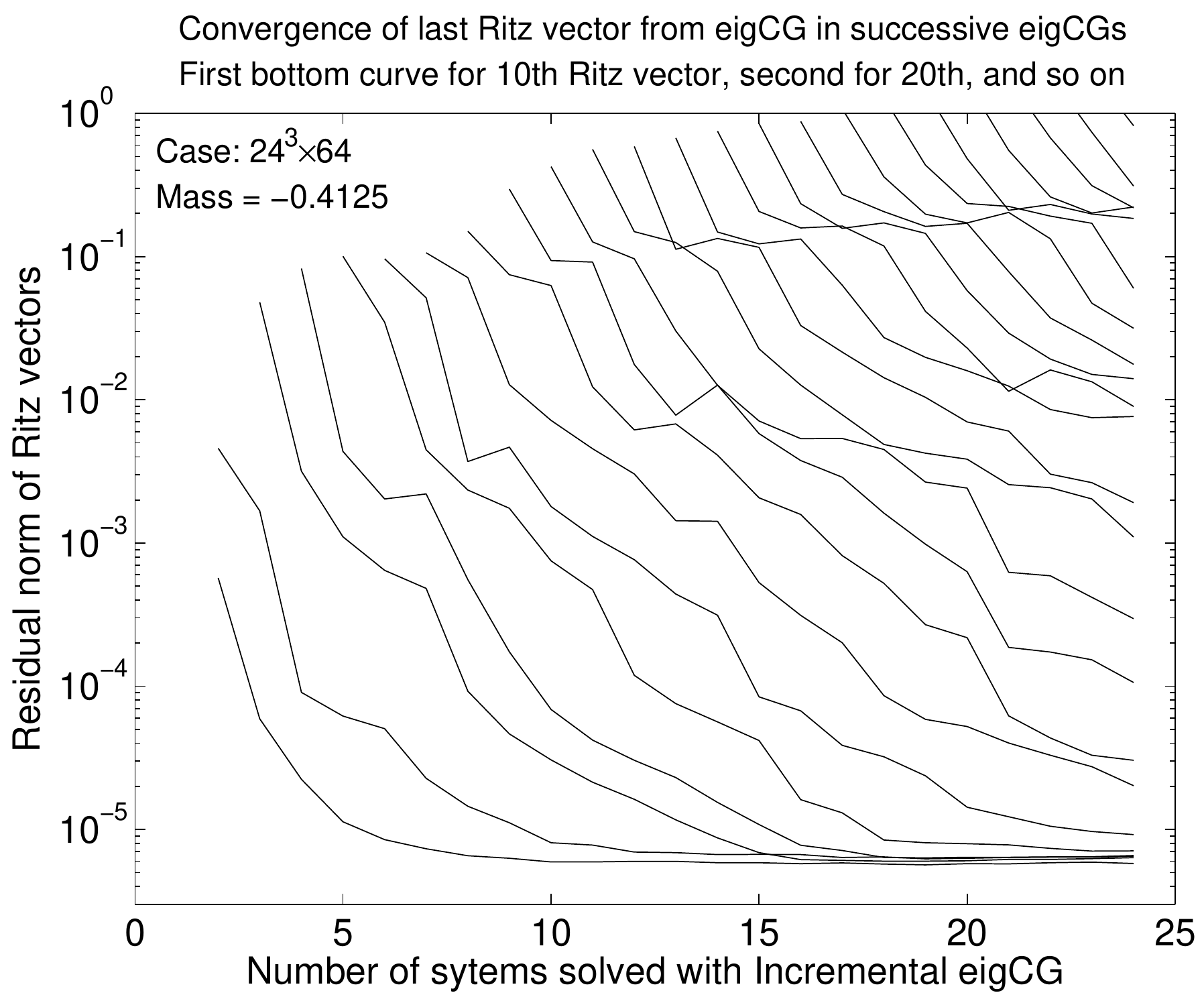}

\includegraphics[width=0.5\textwidth]{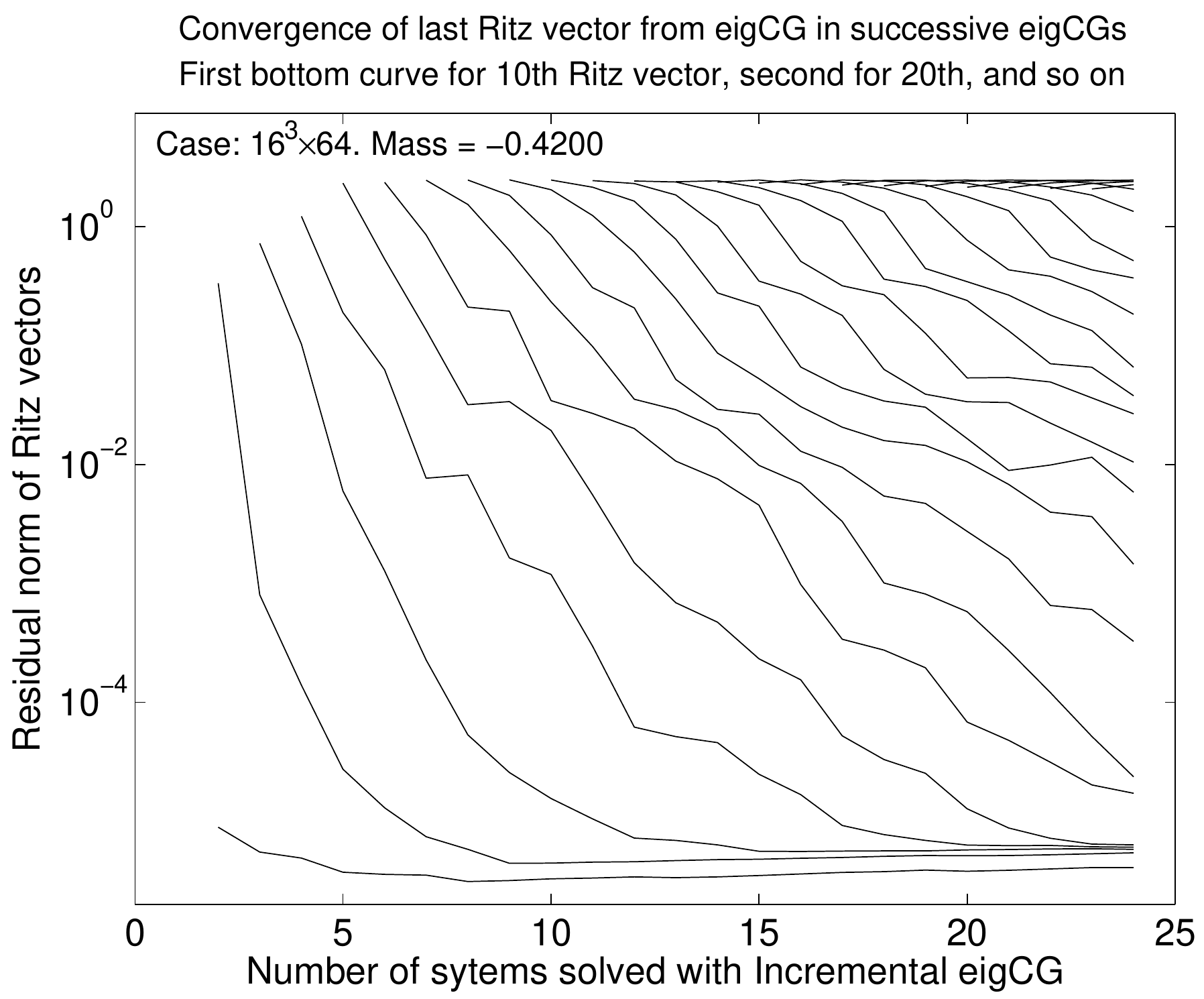}
\includegraphics[width=0.5\textwidth]{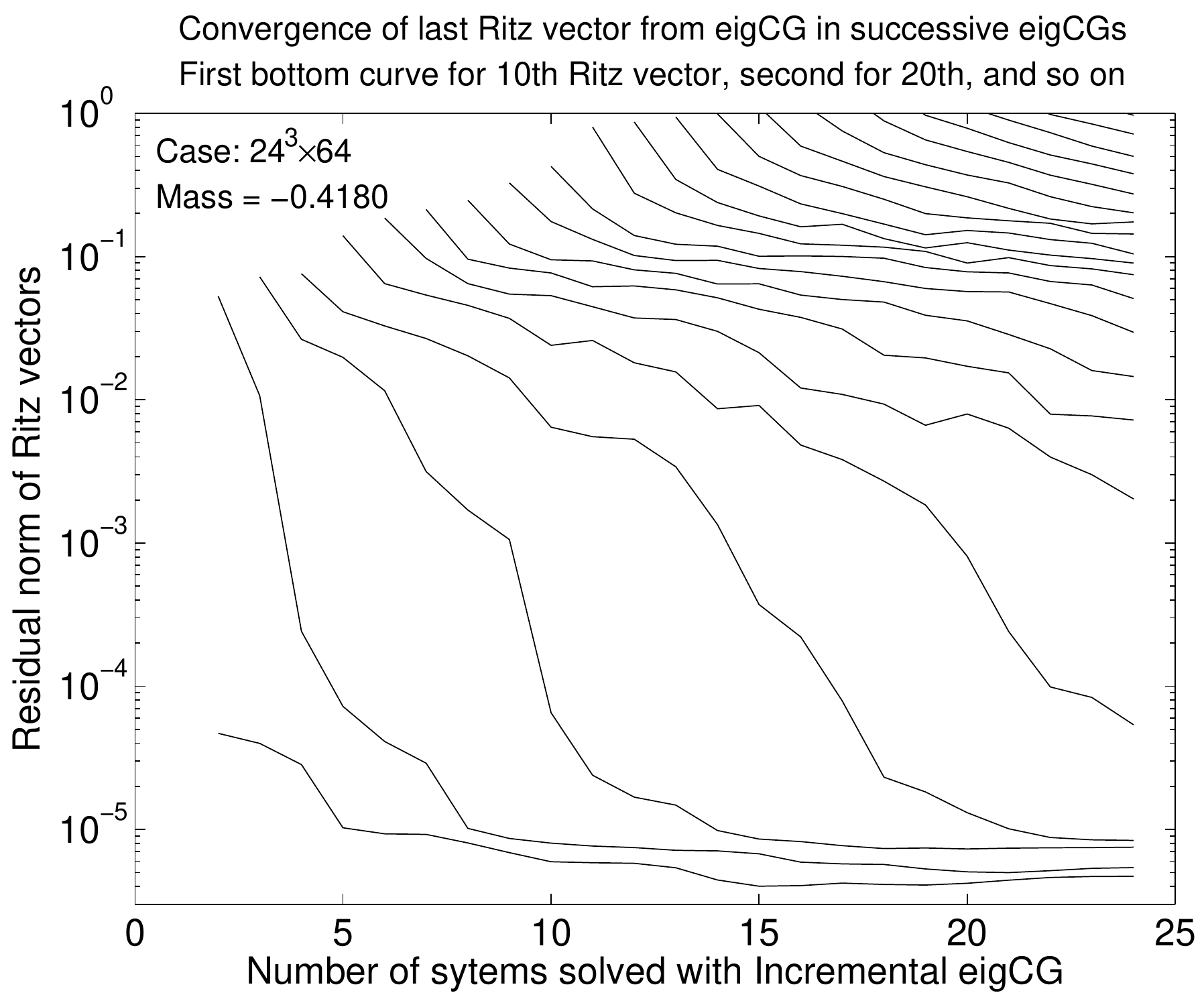}
\caption{Convergence of eigenpairs in the outer loop of 
  Incremental eigCG(10,100). 
A total of 24 linear systems are solved, one per outer iteration.
Each curve shows the convergence of the residual norm of the $(10\times i)$-th
  innermost Ritz vector, which is obtained during the $i$-th outer iteration
  and improved in iterations $i+1, \ldots ,24$.
Typically, residual norms for eigenvalues 
  $(10\times i)+1, \ldots, (10\times i)+9$ fall between the two depicted 
  curves for  $(10\times i)$ and  $(10\times (i+1))$.
Left graphs show results from the 3M lattice and right graphs from 
  the 10M lattice.
Matrices coming from three different masses are considered for each lattice;
  a very heavy mass (top), the sea-quark mass (middle), 
  and a mass close to the critical mass (bottom).
Slower CG convergence with lighter masses allows eigenvalues to be found faster.
}
\label{fig:Incremental eigCG convergence}
\end{figure}

In this section, we show how Incremental eigCG improves partially 
  converged eigenpairs and how it produces additional interior eigenpairs.
Figure \ref{fig:Incremental eigCG convergence} shows the convergence of 
  certain eigenpairs at every outer iteration of the Incremental eigCG
  for each of the two lattices and for three quark masses.
In all cases, we use eigCG(10,100) to solve 24 unrelated right hand sides.
After eigCG converges to a system, the computed 10 Ritz vectors
  augment $U$. 
We explicitly perform a Rayleigh Ritz on $U$ to report the best eigenvector 
  approximations.
We show the convergence of every 10th Ritz pair 
  from the step they were first produced by eigCG and after each outer 
  iteration. 
For example, the convergence history of the 30th smallest eigenpair is the 
  third curve from the bottom in the graphs. 
The curve first appears at outer step 3 and improves after the solution 
  of each subsequent system. 
In all cases, eigenpair approximations continue to converge 
  and more eigenpairs are calculated incrementally with more outer iterations.

The top graphs in Figure \ref{fig:Incremental eigCG convergence}, which 
  correspond to a very heavy mass and thus a small condition number, 
  show that linear systems converge faster than eigenvectors.
Incremental eigCG requires the solution of 10 right hand sides for the 
  3M lattice and 18 right hand sides for the 10M lattice to achieve machine 
  precision for the first 10 eigenpairs (first bottom curve).
One could continue the first eigCG until enough eigenvectors converge, 
  but this would not take advantage of these iterations to solve linear systems.
Moreover, deflating only a couple of eigenvalues is sufficient for 
  low condition numbers.
As the condition number deteriorates with lighter quark masses
  (middle and bottom graphs), eigCG takes far more iterations and 
  thus can obtain the smallest 10 eigenvalues to machine precision
  by solving less than five linear systems.
This behavior is consistent with the distribution of the eigenvalues 
  in Figure \ref{fig:eigenvalues}.

Incremental eigCG finds about the same number of extremal eigenvalues 
  per number of linear systems solved, with
  a gradual decrease in the rate for more interior eigenpairs,
  especially with lighter masses
  (see Figure \ref{fig:Incremental eigCG convergence}). 
The former is expected because eigCG builds a different space for each right 
  hand side, so Incremental eigCG has no information for interior eigenvalues
  until they are targeted.
There are three reasons for the gradual rate decrease.
First, the graphs in the figure are shown against the number of linear
  systems solved so the scale is not representative of the actual work 
  spent to find these eigenvalues.
Indeed, later systems are deflated with more vectors in $U$ and so
  take fewer iterations to solve. 
Second, deflation does not improve the relative gaps between unconverged 
  eigenvalues, so with fewer iterations eigCG cannot recover as much 
  eigen-information by solving one linear system.
Third, as discussed in section \ref{sec:back_initCG} for init-CG,
  the convergence of interior eigenvalues plateaus when it reaches the 
  accuracy of more extreme deflated eigenvalues.
Then, eigCG tries to improve the already computed eigenvectors 
  rather than compute new ones.
Loss of orthogonality in CG can also contribute to this.
In this paper, we do not seek to alleviate this last problem 
  through spectral preconditioning or by orthogonalization as in RMINRES, 
  because the additional cost for obtaining more interior eigenvalues 
  may not be justified as the exterior eigenvalues determine the 
  condition number to a greater extent.
Even so, we show how much can be achieved with no 
  substantial additional cost to CG. 

\begin{figure}
\includegraphics[width=0.5\textwidth]{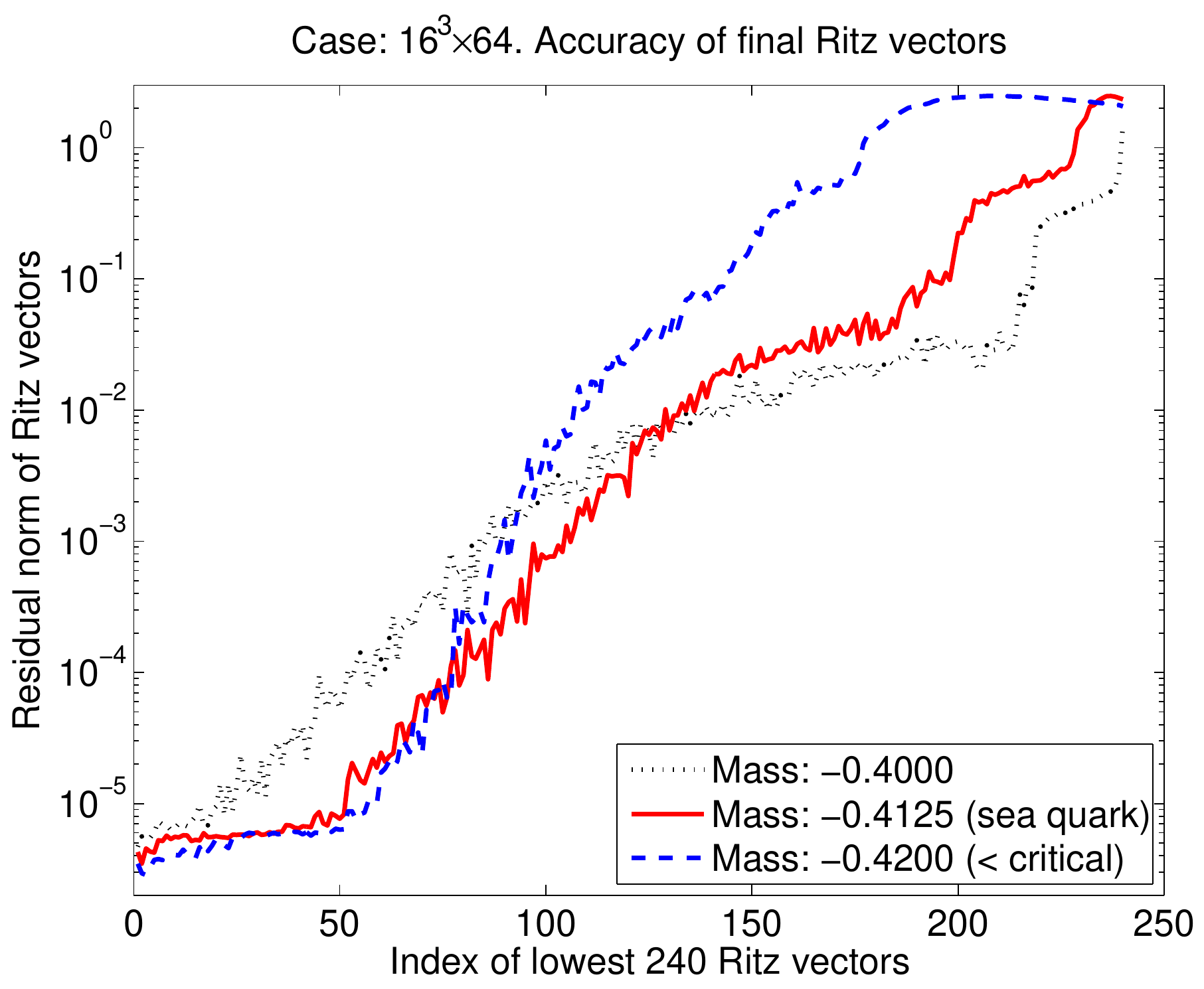}
\includegraphics[width=0.5\textwidth]{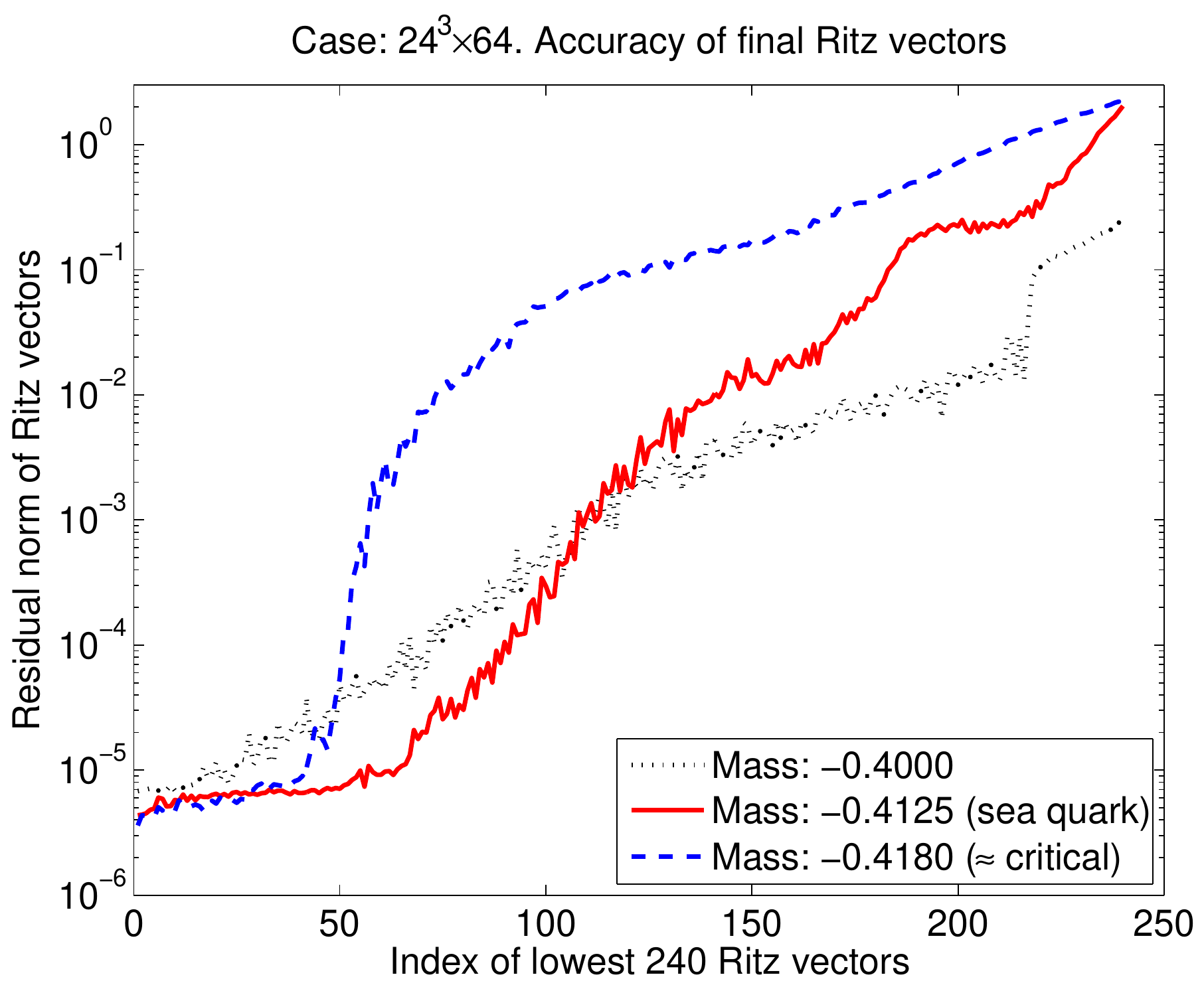}
\caption{Residual norm of the 240 Ritz vectors computed at the end of 
  Incremental eigCG on 24 right hand sides. Left graph shows results from
  the 3M lattice and right graph from the 10M lattice. The three curves 
  correspond to three different quark masses (a heavy, 
  the sea quark, and one close to the critical mass).}
\label{fig:final accuracy of eigenvalues}
\end{figure}

We conclude this study of eigenvalue convergence by showing in 
  Figure \ref{fig:final accuracy of eigenvalues} the residual norms of the
  240 computed Ritz vectors after all 24 linear systems have been
  solved with Incremental eigCG.
The graphs show that for heavy masses more interior eigenvalues are 
  found to better accuracy than with lighter masses, but with lighter 
  masses extreme eigenvalues are found to much better accuracy. 
This is particularly evident in the 10M lattice.
We have found that deflating Ritz vectors with residual norm 
  below $\sqrt{\epsilon_{mach}}$ is more effective.
Thus, about 80-110 eigenpairs can be deflated for both lattices, 
   except for the 10M lattice with a mass close to critical one. 
In that case, a spectrally preconditioned eigCG could further improve
  interior eigenvalues.

\subsection{Linear system convergence with init-CG}
\label{sec:numerical-initcgworks}

\begin{figure}
\includegraphics[width=0.5\textwidth]{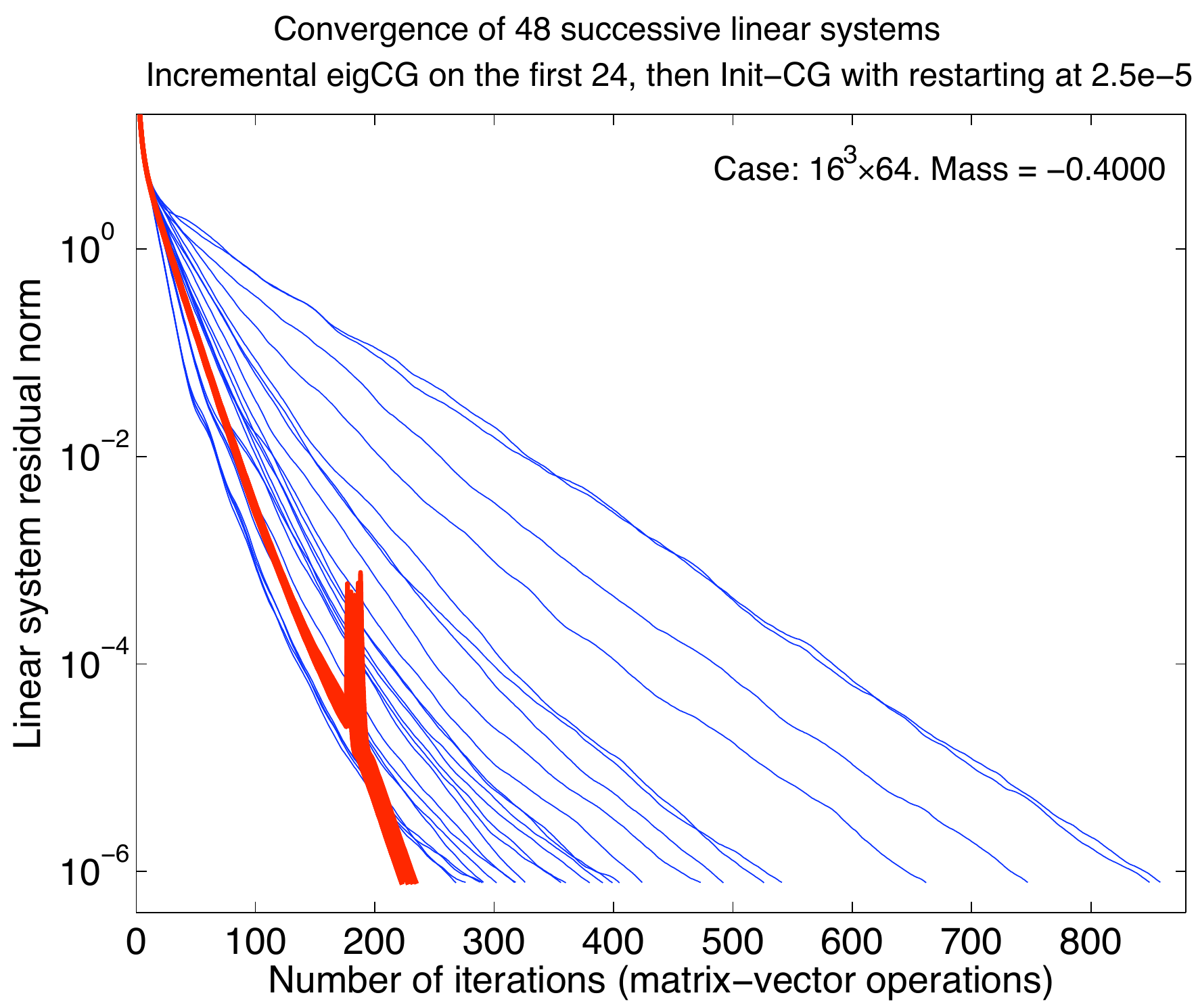}
\includegraphics[width=0.5\textwidth]{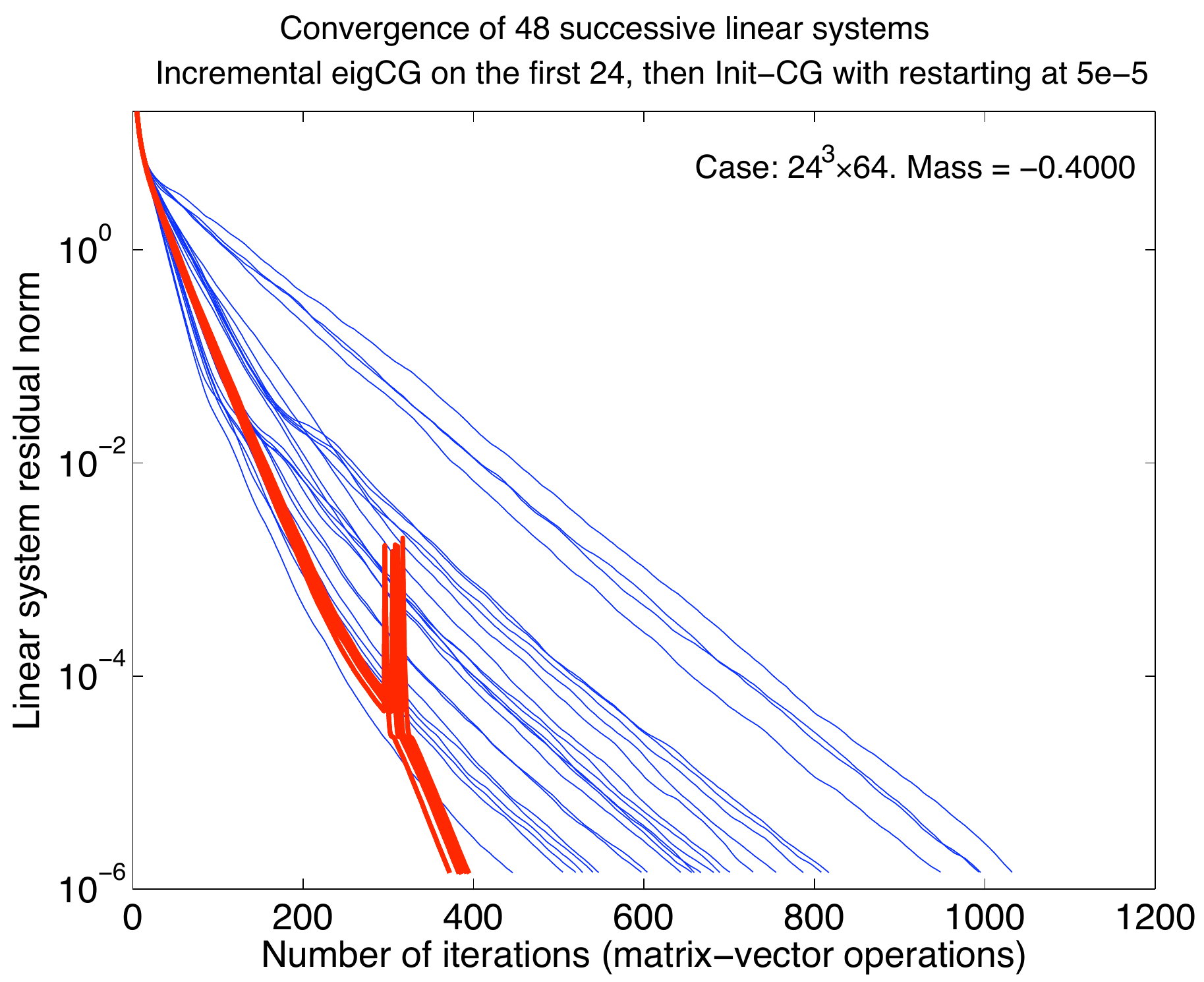}

\includegraphics[width=0.5\textwidth]{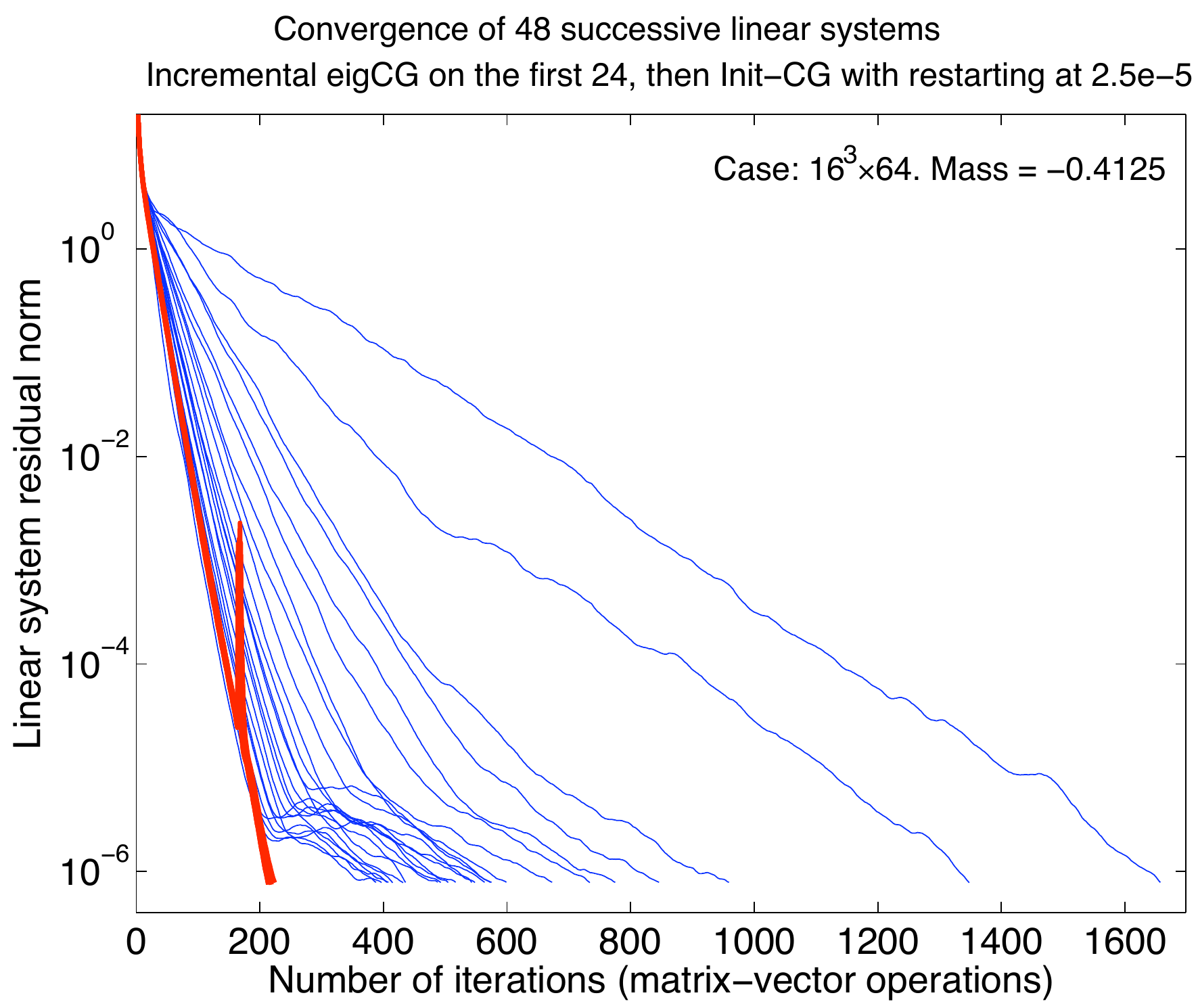}
\includegraphics[width=0.5\textwidth]{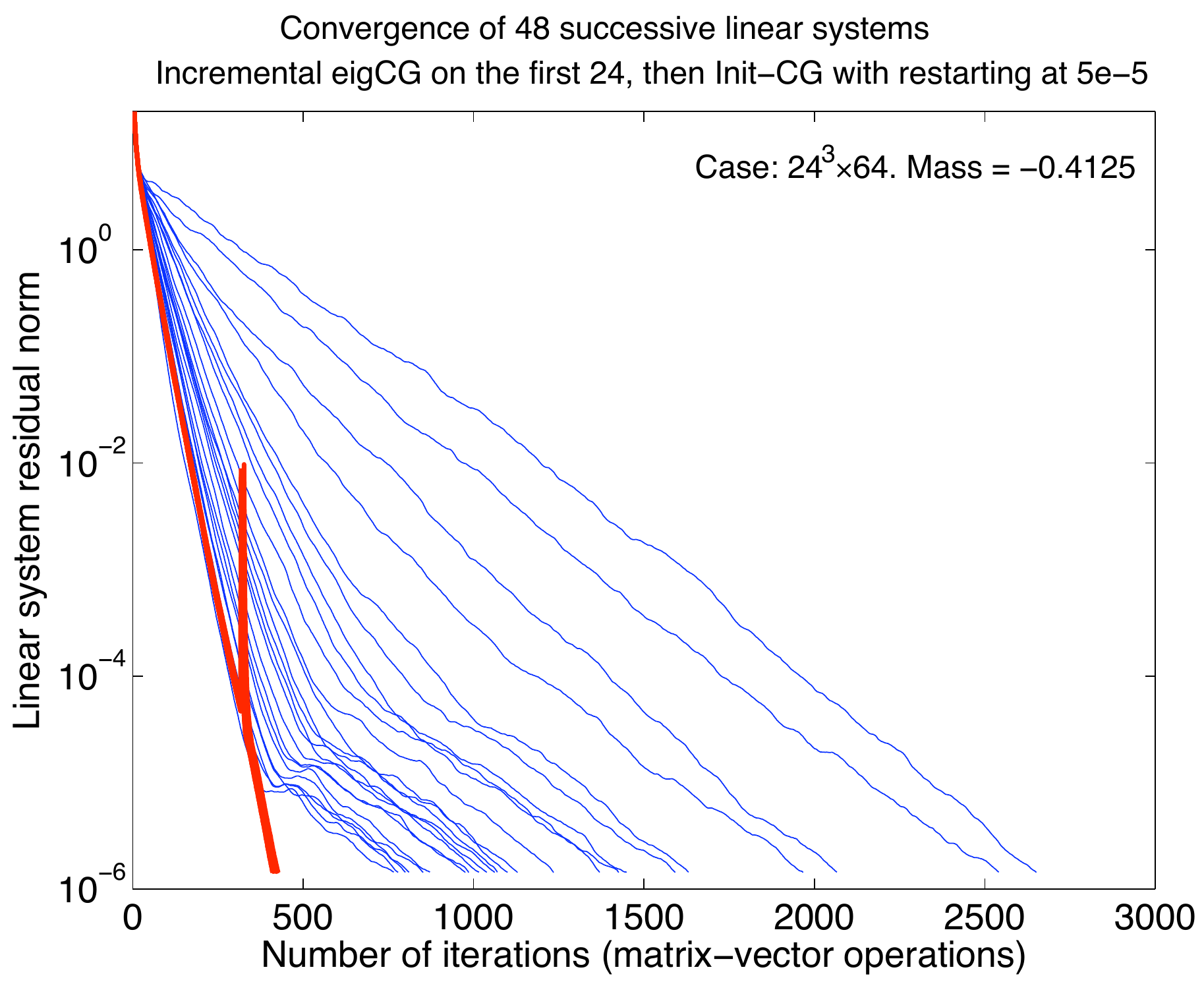}

\includegraphics[width=0.5\textwidth]{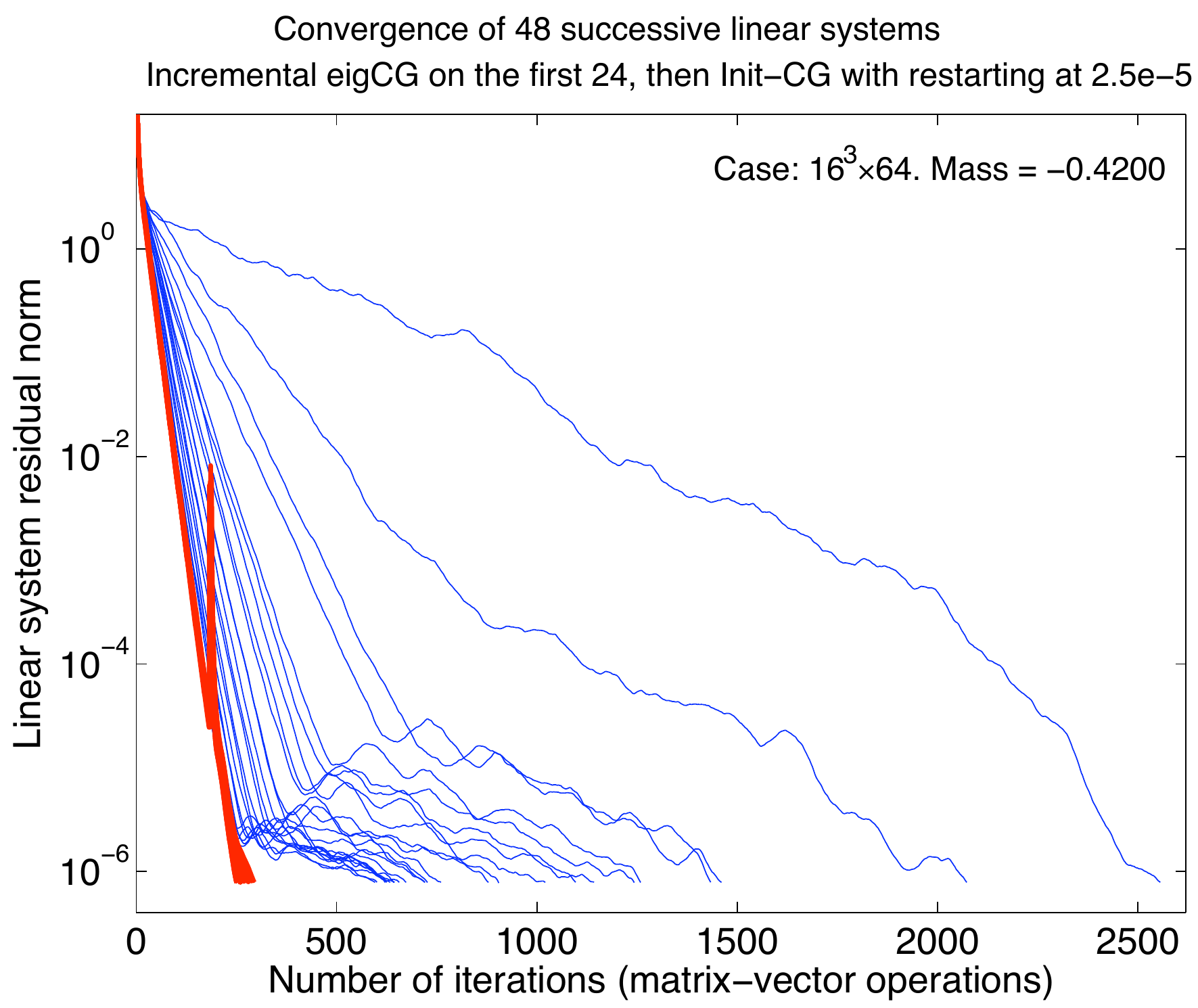}
\includegraphics[width=0.5\textwidth]{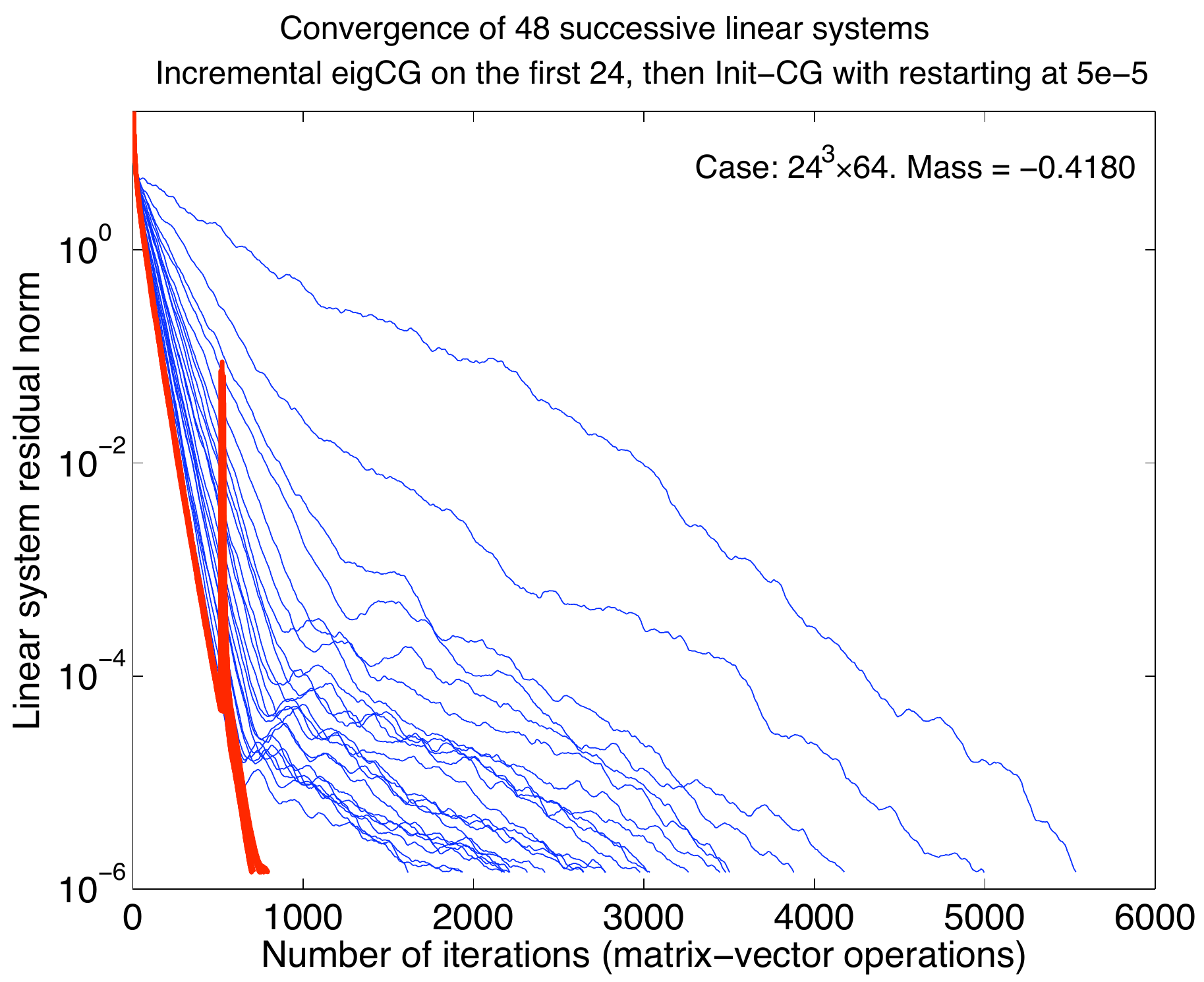}
\caption{Linear system convergence solving 40 unrelated right hand sides. 
Left graphs show results from the 3M lattice and right graphs from 
  the 10M lattice.
Matrices coming from three different masses are considered for each lattice;
  a very heavy mass (top), the sea-quark mass (middle), 
  and a mass close to the critical mass (bottom).
Incremental eigCG(10,100) is used for the first 24 systems.
The final 240 approximate eigenvectors are used in init-CG to solve the
  rest 24 systems.
Restarting CG close to single machine precision resolves any 
  convergence delays associated with init-CG.}
\label{fig:Linear_System_Convergence}
\end{figure}

Figure \ref{fig:Linear_System_Convergence} shows the CG convergence history 
  for solving three linear systems for each of the two lattices, 
  each system having 48 right hand sides. 
For the first 24 right hand sides we use Incremental eigCG(10,100),
  and for the 24 subsequent systems we use init-CG deflated with the 
  obtained 240 approximate eigenvectors.
Therefore, CG convergence is the slowest for the first system without
  deflation and improves as groups of 10 Ritz vectors are accumulated 
  by Incremental eigCG.
As expected from the eigenvalue spectrum of these matrices, there are 
  diminishing returns from deflating an increasing number of eigenvectors.
However, these diminishing returns start approximately at the point where 
  the smallest non-deflated eigenvalue becomes relatively invariant of the 
  quark mass used.
In Figure \ref{fig:Linear_System_Convergence} the init-CG used for the final 
  24 vectors converges in approximately the same number of steps regardless
  of quark mass, yielding speedups of more than 8 in the most difficult cases.

The Incremental eigCG curves for the first 24 systems show a sublinear 
  convergence behavior which begins at the 
  accuracy at which deflated approximate eigenvectors were obtained. 
Instead of a plateau, however, we see a gradual deterioration of the rate
  of convergence because extreme eigenpairs, which have a bigger effect 
  on the condition number, are obtained more accurately than interior ones.
For simplicity, we did not address this problem for the first 24 right hand 
  sides during the Incremental eigCG, but only for the 24 subsequent 
  right hand sides to be solved with init-CG. 
In many applications, including ours, the number of subsequent systems 
  is large so optimizing the initial Incremental eigCG phase may not have 
  a large impact.

For the second phase, we restart the init-CG 
  when the norm of the linear system residual reached within an order of 
  magnitude of machine precision (scaled by the norm of the matrix).
As seen in the graphs of the previous section, most deflation benefits 
  come from Ritz vectors with residual norm below this threshold.
The graphs in Figure \ref{fig:Linear_System_Convergence} show that
  after a short lived peak, the restart completely restores the linear 
  CG convergence.
A dynamic way to choose the restart threshold can be devised based 
  on the computed eigenvalues and their residual norms, and balancing
  the benefits of reducing the condition number with the expense of 
  restarting.
Such a technique goes beyond the scope of this paper.

\subsection{Removing the QCD critical slowdown}
\label{sec:numerical-noslowdown}
\begin{figure}[tp]
\includegraphics[width=0.5\textwidth]{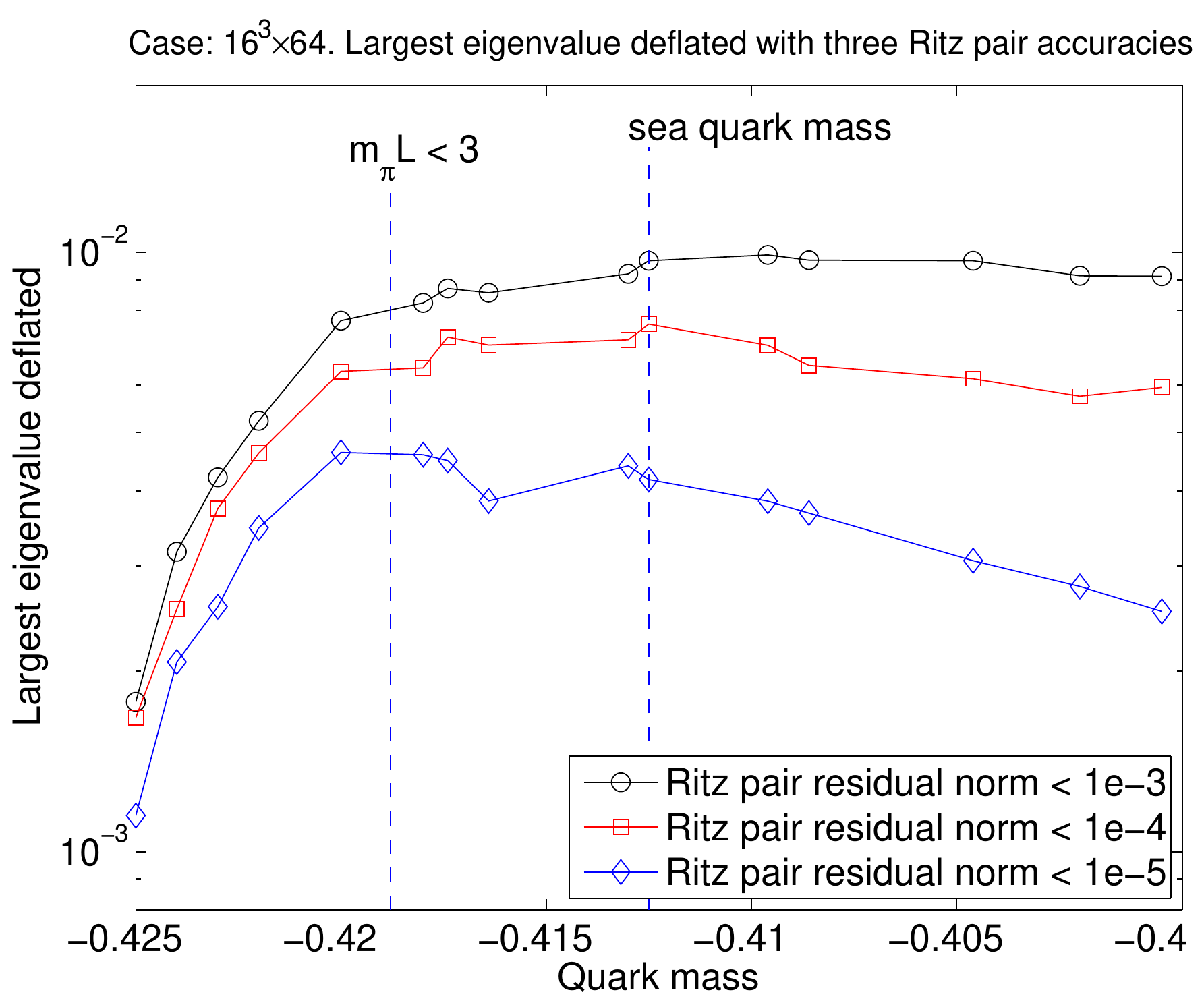}
\includegraphics[width=0.5\textwidth]{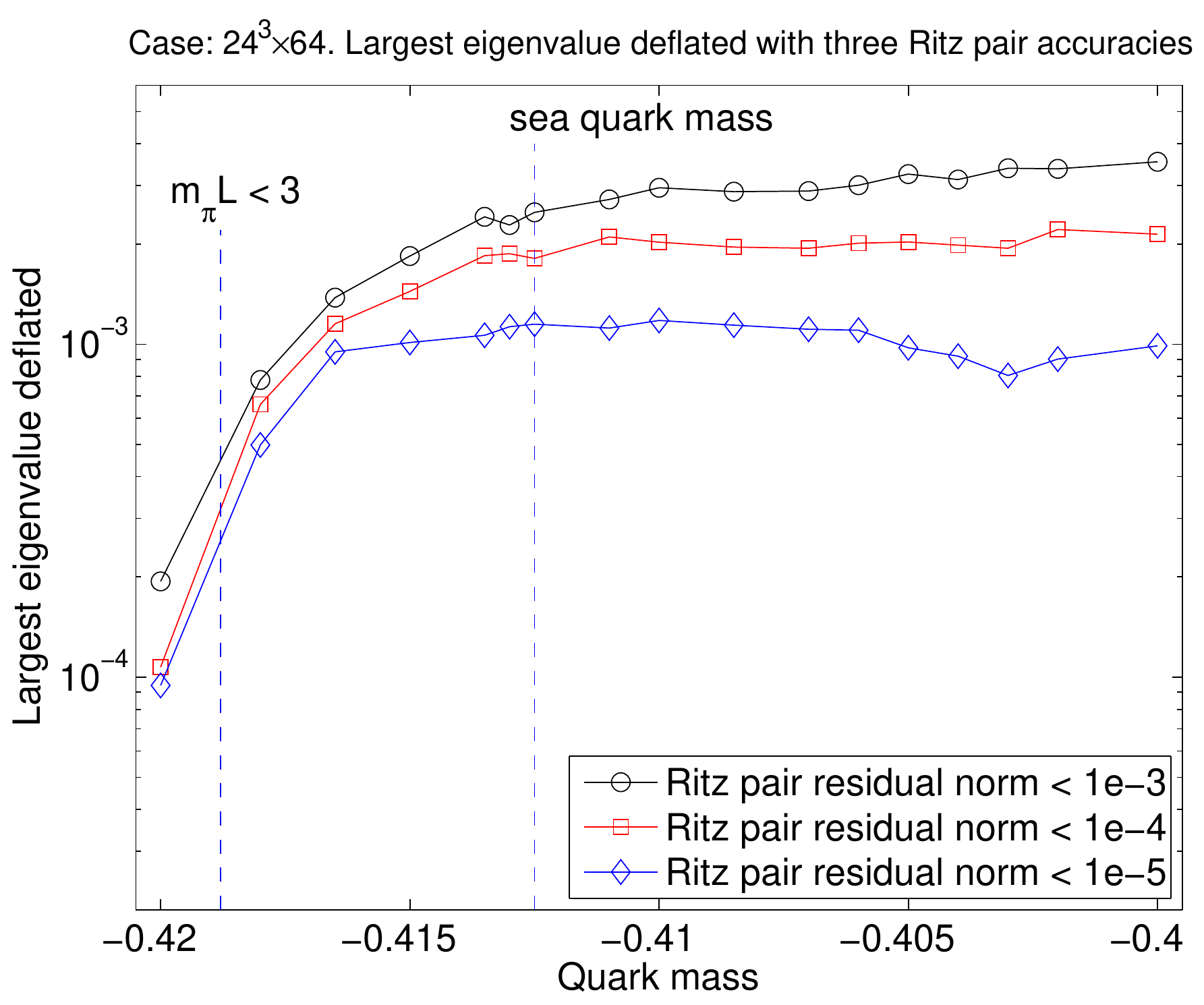}
\caption{For each quark mass and its corresponding matrix we plot the 
  largest eigenvalue of the 240 computed by Incremental eigCG that 
  has residual norm less than some threshold. Each of the three curves
  corresponds to a different threshold. 
}
\label{fig:Eigenvalue no Critical Slowdown}
\end{figure}

\begin{figure}[hp]
\includegraphics[width=0.5\textwidth]{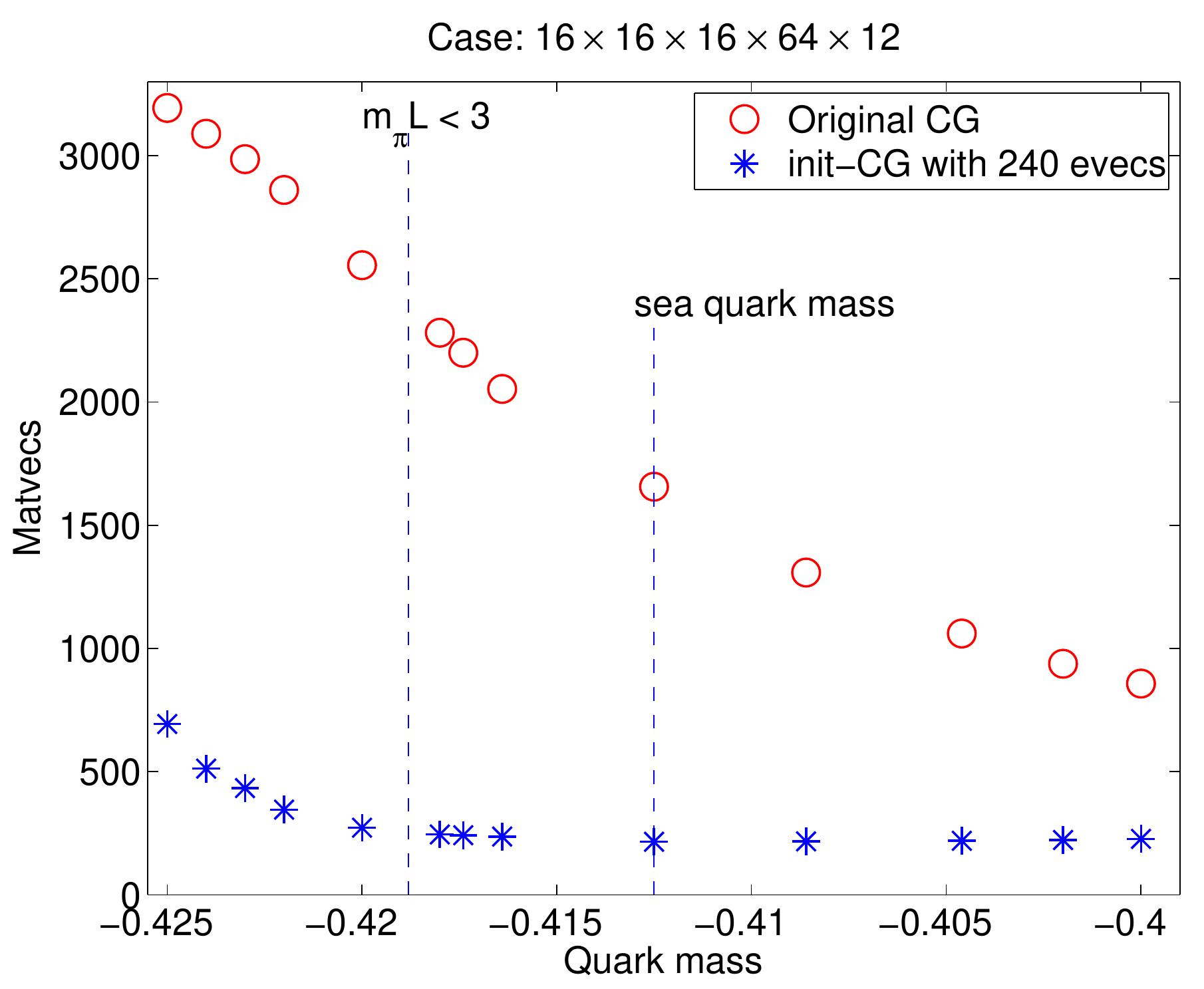}
\includegraphics[width=0.5\textwidth]{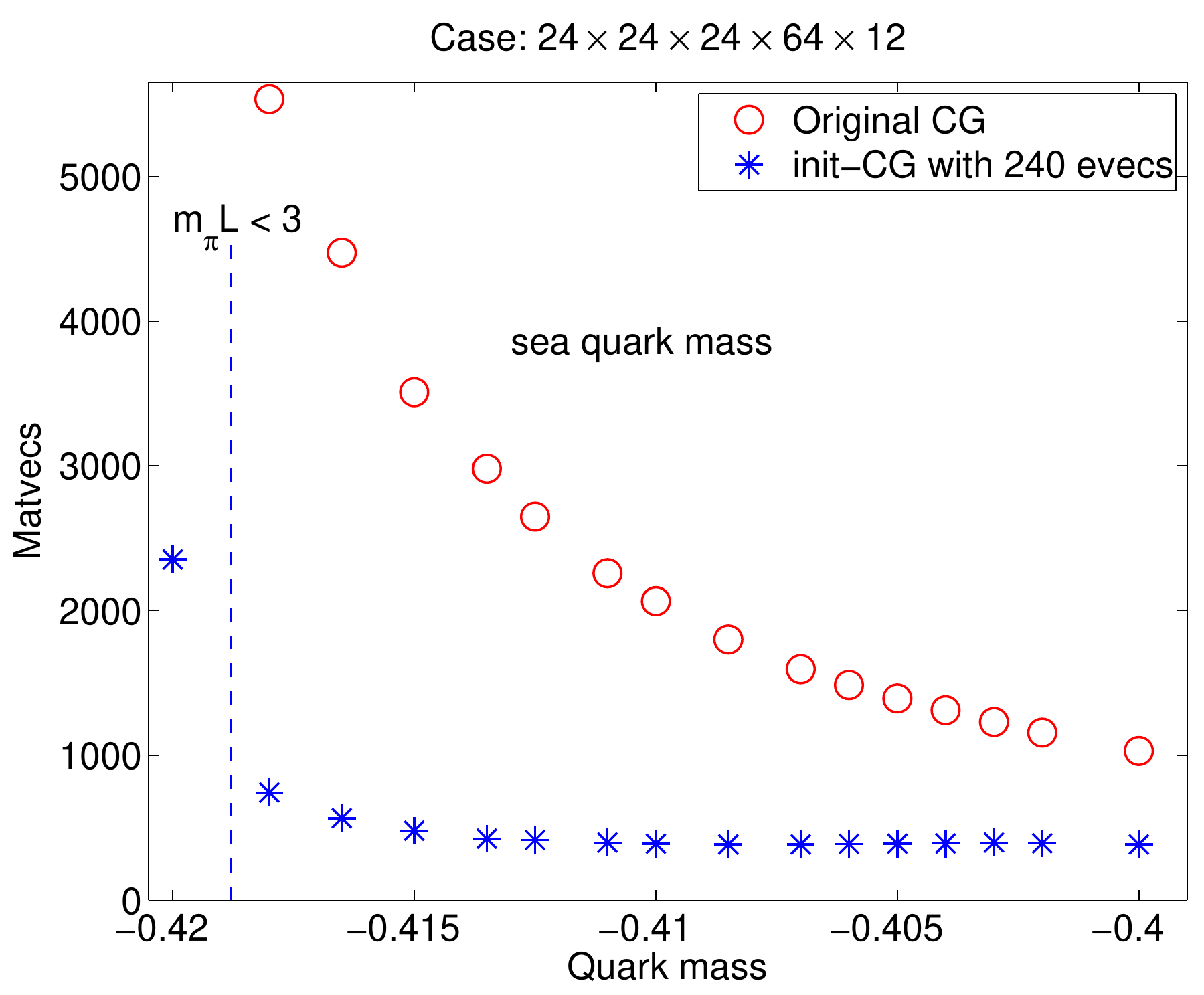}
\caption{For each quark mass and its corresponding matrix we plot the 
  average number of iterations required by init-CG to solve the rest 24 systems.
  For comparison the number of iterations of the non-deflated CG is reported.
}
\label{fig:Iterations no Critical Slowdown}
\end{figure}

We have run similar experiments with 48 right hand sides 
  for both the 3M and 10M lattices on several matrices coming from a 
  wide range of quark masses; from heavy to below critical.
In Figure \ref{fig:Eigenvalue no Critical Slowdown} we show how large 
  an eigenvalue we can expect the init-CG algorithm to deflate and how well.
This eigenvalue is the denominator of the condition number of 
  the deflated operator. 
We consider three thresholds  1E-3, 1E-4, and 1E-5 and plot for each matrix
  the largest eigenvalue returned by Incremental eigCG that has residual
  norm less than these three thresholds. 
For the 3M lattice (left graph) we see that lighter masses allow
  our algorithm to find eigenvalues deeper in the spectrum very accurately. 
For threshold 1E-3, the eigenvalues identified by Incremental eigCG are 
  very close to 0.009 for all physically meaningful masses.
Therefore, we expect similar conditioning and number of iterations 
  regardless of the mass. 
This is confirmed in Figure \ref{fig:Iterations no Critical Slowdown}.
We also note the weakening ability of Incremental eigCG to identify interior 
  eigenvalues below the critical mass because of loss of orthogonality 
  in eigCG/init-CG.
Similar observations can be made for the 10M lattice, with the exception that
  the eigenvalues that are accurate to 1E-3 tend to be smaller near the 
  critical mass.
Still, the ratio between the 1E-3 accurate eigenvalues at masses -0.4112 
  (sea quark mass) and -0.4180 (critical) is less than 4, 
  implying a slowdown of no more than two over heavier masses. 
This is confirmed in Figure \ref{fig:Iterations no Critical Slowdown}.

Figure \ref{fig:Iterations no Critical Slowdown} shows the average
  number of iterations required by init-CG to solve the 24 right hand sides
  for the two lattices and for each mass.
We also plot the number of iterations required by the non-deflated CG. 
Speedups close to an order of magnitude are observed and, more importantly, 
  the number of iterations of init-CG is almost constant across meaningful 
  masses.
Again, we note that a more fastidious use of spectrally preconditioned eigCG 
  would have resulted in further reduction in iterations, 
  especially for the 10M lattice, but this reduction would have been far
  less substantial relative to those reported in this paper. 
Moreover, this would only be needed in 
  the physically non-meaningful range of masses (3M lattice) or very 
  close to the critical mass (10M lattice).
Instead, we showed why the critical slowdown can be removed in principle 
  when the number of right hand sides is large and derived an 
  algorithm that achieves this.

\subsection{Cray XT4 timings}
\label{sec:numerical-timings}

We have run the 10M lattice on the Cray XT4 at NERSC for a real world, 
  all-to-all propagator calculation~\cite{Foley:2005ac} 
  with time and spatial even-odd dilution. 
Two different random noise vectors were used for a total of 
  256 right hand sides. 
This is a stochastic method of estimating all matrix elements of the 
  quark propagator (i.e. the Dirac matrix inverse). 
For details regarding all-to-all propagator calculations 
  see~\cite{Kamleh:2005wg}.
The quark mass used was $m_{q} = -0.4125$ which is the same as the 
  dynamical quark mass used to generate the gauge configurations. 
The corresponding pion mass was determined to be roughly 400MeV and the 
  spatial box was about 2.6fm. 
These are typical parameters in current lattice QCD calculations, 
  although lighter masses would clearly benefit our methods.
Our codes are compiled with C++ using the -O2 and loop unrolling flags.

As in our previous experiments we run two phases: First we
  apply Incremental eigCG(10,100) on the first 24 right hand sides. 
Second, on the following 232 systems, we apply init-CG deflated by 
  the accumulated 240 vectors in $U$.
We also restart the init-CG as in the previous sections, thereby incurring 
  the deflation cost twice.

Table \ref{tab:timings} shows the execution times for the two phases 
  and the overall time for the application.
We also report times for the original CG code as implemented in Chroma. 
All codes use the Chroma implementation of the sparse matrix-vector 
  multiplication.
The native CG in Chroma implements various architectural optimizations 
  such as SSE vector processing and hand loop unrolling.
Our CG implementation achieves an iteration cost only about 2\% more 
  expensive than the native Chroma CG.
However, the benefits of deflation are far more significant.
Linear systems deflated with the 240 vectors are solved 5.4 times faster
  than regular CG. 
Moreover, while obtaining these 240 vectors, our algorithm is still  
  1.6 times faster than if we were to simply run CG. 
Overall, the application runs with a speedup of 4.2.

\begin{table}
\caption{Execution times in seconds for phase 1 where we apply 
  Incremental eigCG, for phase 2 where we switch to init-CG,
  and for the whole application that includes also about 90 seconds of 
  application setup time. We also show the times from using the original CG 
  in Chroma without deflation.}

\label{tab:timings}
\begin{center}
\begin{tabular}{|rr|rr|rr|}
\hline
\multicolumn{2}{|c|}{Time for first 24 rhs} 
	 &  \multicolumn{2}{c|}{Time for next 232 rhs} & 
			  \multicolumn{2}{c|}{Total application time} \\
\hline
 Chroma CG   & 527.9 & Chroma CG & 5127.2  & Chroma CG 	   & 5751.5 \\
Incr. eigCG  & 323.2 & init-CG   & 951.2   & Incr. eigCG   & 1365.8 \\
\hline
     speedup & 1.6  & speedup  & 5.4     & overall speedup &  4.2 \\
\hline
\end{tabular}
\end{center}
\end{table}

\begin{figure}[t]
\includegraphics[width=0.5\textwidth]{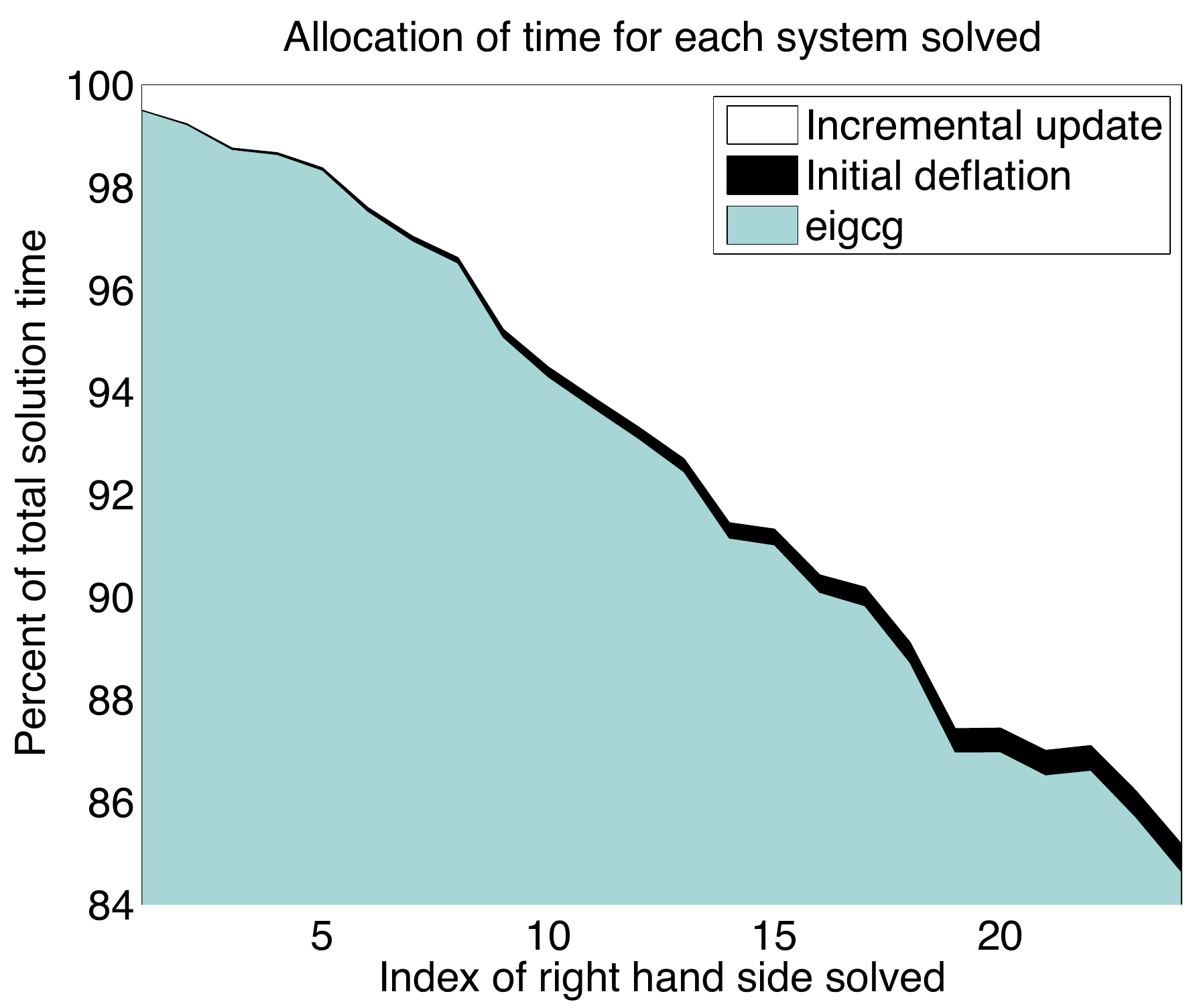}
\includegraphics[width=0.5\textwidth]{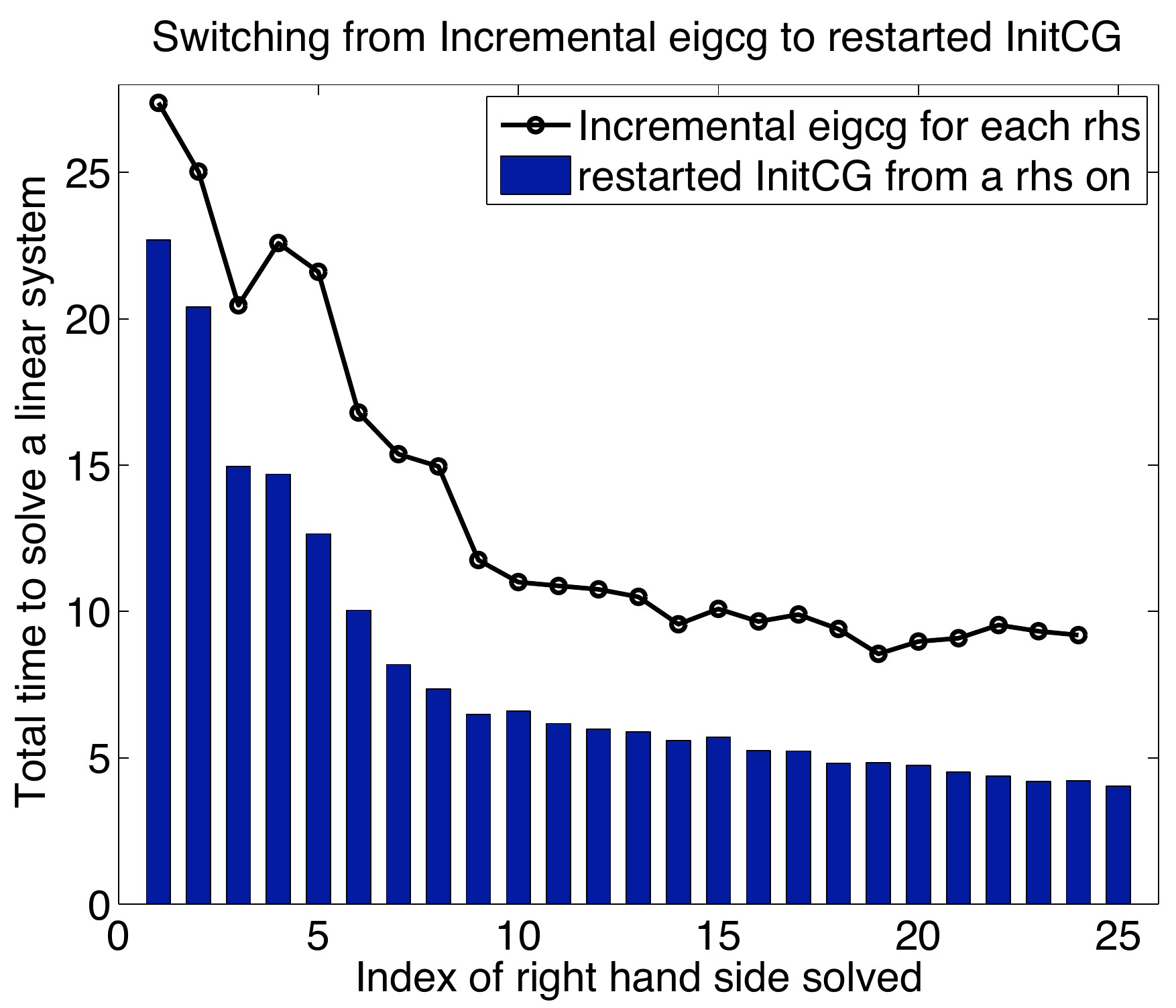}
\caption{
Left graph: Percentage breakdown of execution time for the three
  components of Incremental eigCG:
  the eigCG inner iteration, the initial deflation, and the incremental
  update.
Right graph:
  The solid line shows the execution time in seconds of Incremental eigCG 
  for the $k$-th right hand side, $k=1,\ldots ,24$.
  The bars below each point show the time it would take to solve a 
  linear system by switching to init-CG at the $k$-th right hand side.
}
\label{fig:timings}
\end{figure}

Next, we quantify the relative expense of the various components in our
  algorithm; specifically, the relative costs
  of deflating with $U$, 
  of incrementally updating $U$ outside eigCG,
  and of updating $V$ inside eigCG.
In Figure \ref{fig:timings}, the left graph shows the percentage breakdown
  of execution time among eigCG, the incremental update, 
  and the deflation part of init-CG.
We show these only for the 24 vectors in the first phase where Incremental
  eigCG is used.

Our first observation is that the initial deflation part is negligible. 
Even deflating 230 vectors (before solving the 24th right hand side) 
  constitutes less than 0.5\% of the time spent in Incremental eigCG.
In the second phase, when deflation with 240 vectors occurs twice and when 
  the cheaper CG (i.e., eigCG(0,0)) is used, the total expense of deflation 
  is less than 3\% of the time to solve a system.

The cost of the incremental update of $U$ increases linearly with the 
  number of right hand sides.
As shown in Figure \ref{fig:timings}, the cost of updating 230 vectors with 
  10 additional ones (the 24th step) constitutes about 15\% of the total 
  Incremental eigCG time. 
This is a result of re-orthogonalizing the new vectors $V$ against $U$ 
  and also of the fact that the Incremental eigCG is about 2.5 times faster 
  than the original eigCG, so the relative cost of updates is pronounced.
Lastly, in our application, updating $V$ during eigCG(10,100) costs an 
  additional 21\% over simple CG.
Clearly, these relative costs depend on the cost of the matrix vector 
  operation, and, for the general case, one must refer to the computational 
  models in Sections \ref{sec:ourmethod} and \ref{sec:incremental}.

Finally, we study the scenario where Incremental eigCG is run on fewer 
  right hand sides, hence accumulating fewer deflation vectors for phase two.
The right graph in Figure \ref{fig:timings} shows two curves.
The points on the solid line show the time taken by Incremental eigCG for
  the $k$-th right hand side, $k=1,\ldots ,24$. 
Immediately below each point, the bar represents the time init-CG would 
  take to solve a linear system deflating only the $10(k-1)$ vectors 
  accumulated by Incremental eigCG up to the $(k-1)$-th right hand side.
The lower runtime is due to three reasons: We avoid the eigCG-related costs, 
  the cost of updating $U$, and last but most important, 
  the restarting of init-CG avoids the plateaus shown in 
  Figure~\ref{fig:Linear_System_Convergence}, speeding the method by 
  a factor of two (when $k =24$).

Figure \ref{fig:timings} also shows that the most significant speedups are 
  obtained by running Incremental eigCG for 8-9 right hand sides 
  (deflating 80-90 vectors).
After that improvements wane but still manage to add an additional factor 
  of 1.6 by $k=24$.
It is possible to monitor these improvements at runtime, thus
  running Incremental eigCG for the number of right hand sides that will 
  minimize overall execution time, before switching to init-CG. 
For example, the following table presents the optimal $k$ for this 
  problem under various numbers of total right hand sides:
\begin{center}
\begin{tabular}{|c|ccccc|}
\hline
total number of RHS		    & 6 & 12 & 24 & 32 & 48 \\
optimal \# RHS in Incremental eigCG & 3 & 7  &  9 & 11 & 19 \\ 
\hline
\end{tabular}
\end{center}
Clearly, the more right hand sides we need to solve, the more 
  vectors it pays to solve with Incremental eigCG.

\section{Conclusions}
The numerical solution of large linear systems with multiple right hand sides
  is becoming increasingly important in many applications.
Our original goal was to address this problem in the context of lattice QCD
  where, in certain problems, hundreds of linear systems of equations must
  be solved.
For general SPD matrices, we have argued that extreme invariant subspaces 
  are the only useful information that can be shared between Krylov spaces 
  built by different, unrelated initial vectors. 
We have also argued that the critical slowdown, observed in lattice QCD
  computations when the quark mass approaches a critical value, is caused 
  by a decrease in exactly the same extreme (smallest) eigenvalues, while the
  average density of more interior eigenvalues remains unaffected.
In our approach we take advantage of the many right hand sides by
  incrementally building eigenspace information while solving linear systems.
This eigenspace is used to accelerate by deflation subsequent linear systems 
  and thus remove the critical slowdown.

The algorithm we have developed that derives eigenspace information during 
  the CG method distinguishes itself from other deflation methods in 
  several ways.
First, we do not use restarted methods, such as GMRES($m$), so our linear
  system solver maintains the optimal convergence of CG. 
Second, by using the readily available CG iterates, we build a local 
  window of Lanczos vectors with minimal additional expense.
Third, we use the locally optimal restarting technique to keep the size 
  of the window bounded. 
Our resulting algorithm, eigCG, has the remarkable property that the
  Ritz pairs converge identically to the unrestarted Lanczos method, 
  to very good accuracy and without having to store the Lanczos vectors.
In our experiments, we were able to find 50-80 eigenpairs to machine precision
  by solving 24 linear systems. 
Current state-of-the-art eigenvalue eigensolvers would require the equivalent
  of 50-80 linear system solves to produce the same information.

We believe the proposed eigCG is a breakthrough method. 
Because it is purely algebraic, it goes beyond lattice QCD to any SPD 
  problem with multiple right hand sides.
Moreover, it does not require the right hand sides to be available 
  at the same time, so it is ideal for time dependent problems.
In this paper, we have left some questions unanswered (especially those
  relating to the theoretical understanding of the method) and pointed to
  many directions that eigCG can be improved. 
Among this wealth of future research, a particularly exciting direction
  is a new eigensolver that redefines the state-of-the-art in the area.
Finally, a general purpose code is currently under development.


\end{document}